\documentclass[english]{lni}

\pdfoutput=1

\usepackage{subcaption}

\usepackage[]{todonotes}

\usepackage{fontawesome}

\usepackage{multirow}

\usepackage{paralist}

\usepackage{amsthm}

\usepackage{threeparttable}

\usepackage[capitalise,nameinlink]{cleveref}
\crefname{section}{Sect.}{Sect.}
\Crefname{section}{Section}{Sections}
\crefname{figure}{Fig.}{Figs.}
\Crefname{figure}{Figure}{Figures}
\crefname{table}{Tab.}{Tabs.}
\Crefname{table}{Table}{Tables}

\usepackage{tikz}
\usepackage{xcolor}
\newcommand*\circled[1]{\tikz[baseline=(char.base)]{\node[shape=circle,fill,inner sep=.5pt] (char) {\footnotesize\textcolor{white}{#1}};}}

\newcommand{\eg}{e.\,g.,\ }
\newcommand{\ie}{i.\,e.,\ }

\newtheorem*{hypothesis}{Hypothesis}
\newtheorem{example}{Example}

\newcommand{\labeltitle}[1]{\vskip -0.5em \noindent\textbf{#1}} 

\begin{document}

\title{Exploring Memory Access Patterns for Graph Processing Accelerators}

\author[Jonas Dann \and Daniel Ritter \and Holger Fröning]
{Jonas Dann\footnote{SAP SE, \email{{firstname.lastname}@sap.com}} \and
Daniel Ritter\footnotemark[1] \and
Holger Fröning\footnote{Heidelberg University, holger.froening@ziti.uni-heidelberg.de}}

\startpage{1}
\editor{Publisher et al.}
\booktitle{Conference}
\year{2021}
\maketitle

\begin{abstract}
Recent trends in business and technology (\eg machine learning, social network analysis) benefit from storing and processing growing amounts of graph-structured data in databases and data science platforms.
FPGAs as accelerators for graph processing with a customizable memory hierarchy promise solving performance problems caused by inherent irregular memory access patterns on traditional hardware (\eg CPU).
However, developing such hardware accelerators is yet time-consuming and difficult and benchmarking is non-standardized, hindering comprehension of the impact of memory access pattern changes and systematic engineering of graph processing accelerators.

In this work, we propose a simulation environment for the analysis of graph processing accelerators based on simulating their memory access patterns.
Further, we evaluate our approach on two state-of-the-art FPGA graph processing accelerators and show \emph{reproducibility}, \emph{comparablity}, as well as the shortened development process by an example.
Not implementing the cycle-accurate internal data flow on accelerator hardware like FPGAs significantly reduces the implementation time, increases the benchmark parameter transparency, and allows comparison of graph processing approaches.
\end{abstract}

\begin{keywords}
DRAM \and FPGA \and Graph processing \and Irregular memory access patterns \and Simulation 
\end{keywords}

\vspace{-0.3cm}
\section{Introduction}
\label{sec:intro}
\vspace{-0.1cm}
Recently, areas in computer science like machine learning, computational sciences, medical applications, and social network analysis drove a trend to represent, store, and process structured data as graphs \cite{journals/corr/abs-1910-09017, journals/corr/abs-2007-07595}.
Consequently, graph processing gained relevance in the fields of non-relational databases and analytics platforms.
As a possible solution to the performance problems on traditional hardware (\eg CPUs) caused by irregular memory accesses and little computational intensity inherent to graph processing \cite{journals/corr/abs-1910-09017, journals/corr/abs-2007-07595, LumsdaineGHB07}, FPGA accelerators emerged to enable unique memory access pattern and control flow optimizations \cite{journals/corr/abs-2007-07595}.
FPGAs, compared to CPUs or GPUs with their fixed memory hierarchy, have custom-usable on-chip memory and logic resources that are not constrained to a predefined architecture.
\Cref{ex:motivation} illustrates the effect of irregular memory accesses for breadth-first search (BFS) with an edge-centric approach.
When not reading sequentially from DRAM, bandwidth degrades quickly \cite{Drepper07}, due to significant latency introduced by DRAM row switching and partially discarded fetched cache lines.
\begin{example} \label{ex:motivation}
    Let each cache line consist of two values, the current BFS iteration be $1$ with root $v_0$, and $e_2$ be the current edge to be processed.
    \Cref{fig:problem} shows an example graph with a simplified representation in DRAM memory.
    The graphs edge array is stored in rows $r_0$--$r_4$ and the current value array is stored in $r_5$ and $r_6$.
    We begin by reading edge $e_2$ which incurs activating $r_1$ in the memory and reading a full cache line.
    Then, we activate $r_5$ and read the first cache line containing $v_0$ and $v_1$, but only use $v_0$.
    Finally, we activate $r_6$ to read $v_5$ and write the new value $1$ to the same location, while wasting bandwidth of one value on each request (\ie reading and not writing the value of $v_4$ respectively).
    \begin{figure}[t]
	    \centering
	    \includegraphics[width=.75\linewidth]{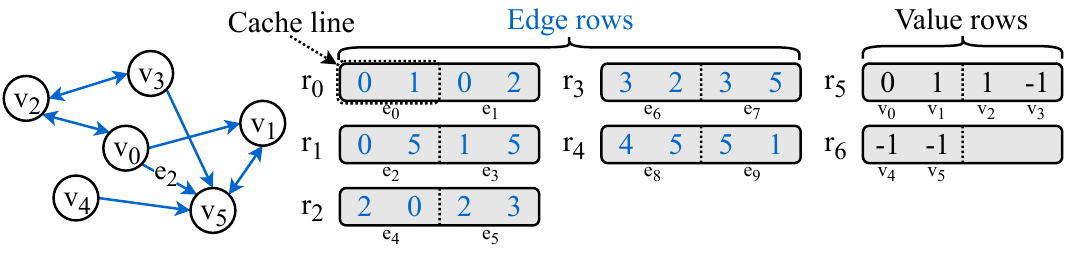}
	    \caption{Illustration of irregular memory accesses for breadth-first search}
	    \label{fig:problem}
    \end{figure}
\end{example}
\vspace{-0.3cm}

While FPGA-based graph processing accelerators show good results for irregular memory access pattern acceleration (\eg \cite{conf/IEEEpact/Yao0L0H18, journals/tpds/ZhouKPSW19}), programming FPGAs is time-consuming and difficult compared to CPUs and GPUs where the software stack is much better developed \cite{journals/sigmod/AbadiAABBBBCCDD19, journals/queue/BaconRS13}.
Additionally, software developers currently lack the skill-set needed for high-performance FPGA programming, making development even more cumbersome.
Aside from that, there are deficiencies in benchmarking of graph processing accelerators due to a large number of FPGAs on the market (almost every article uses a different FPGA), but also lack of accepted benchmark standards (cf. \cite{journals/corr/abs-2007-07595}).
This leads us to the two main challenges in the field:
\begin{inparaenum}[\it (1)] 
    \item time-consuming and difficult development of accelerators for irregular memory access patterns of graph processing,
    \item differences in hardware platforms and benchmark setups hindering reproduction and comparison.
\end{inparaenum}

\begin{figure}[bt]
\centering
\captionsetup[subfigure]{position=b}
\subcaptionbox{Memory access simulation idea\label{fig:idea}} {\includegraphics[width=.45\linewidth]{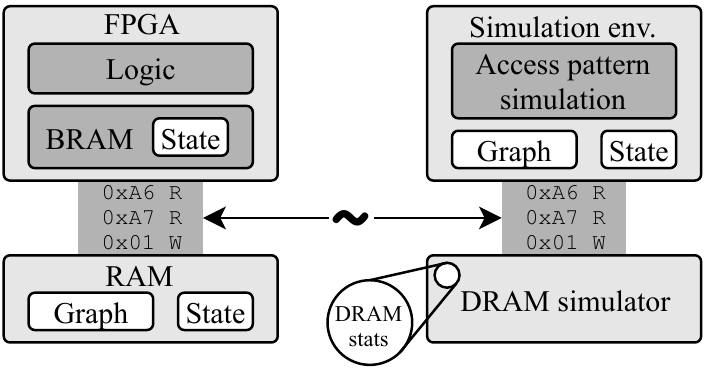}}
\subcaptionbox{Reproducibility error from simulation\label{fig:error}} {\includegraphics[width=.5\linewidth]{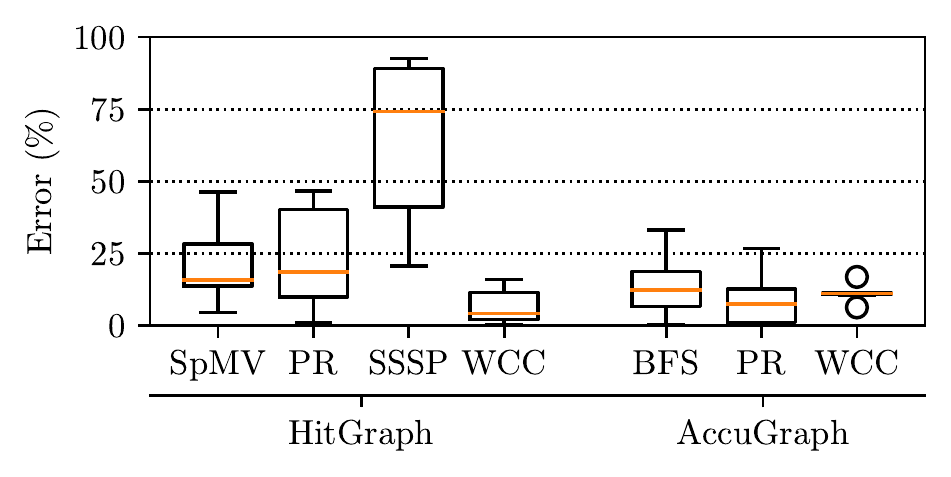}}
\caption{Graph processing memory simulation idea and achieved reproducibility error}
\end{figure}

To solve challenges \emph{(1)} and \emph{(2)}, we propose a simulation environment for graph processing accelerators (based on the idea in \cref{fig:idea}) as a methodology and tool to quickly reproduce and compare different approaches in a synthetic, fixed environment.
On a real FPGA, the on-chip logic implements data flow on on-chip (in block RAM (BRAM)) and off-chip state and graph data in the off-chip DRAM.
Based on the observation that the access to DRAM is the dominating factor in graph processing, we however only implement an approximation of the off-chip memory access pattern in our environment working on the graph and state independently of the concrete (difficult to implement) data flow on the FPGA and feed that into a DRAM simulator.
While the performance reported by such a simulation may not perfectly match real performance measurements, we see a high potential to better understand graph processing accelerators.
This results in the following hypothesis:
\begin{hypothesis}
Memory access patterns dominate the overall runtime of graph processing such that disregarding the internal data flow results in a reasonable error of a simulation.
\label{hyp:1}
\end{hypothesis}

\vspace{-0.3cm}
Our simulation approach significantly reduces the time to test new graph processing accelerator ideas and also enables design support and deeper inspection with DRAM statistics as well as easy parameter variation.
In a recent survey \cite{journals/corr/abs-2007-07595}, we found multiple graph processing accelerator approaches (\eg AccuGraph \cite{conf/IEEEpact/Yao0L0H18}, ForeGraph \cite{conf/fpga/DaiHCXWY17}, HitGraph \cite{journals/tpds/ZhouKPSW19}, and Zhang et al. \cite{conf/fpga/ZhangL18} among others).
Based on criteria like reported performance numbers on commodity hardware and sufficient conceptual details, we chose two state-of-the-art systems -- namely HitGraph \cite{journals/tpds/ZhouKPSW19} and AccuGraph \cite{conf/IEEEpact/Yao0L0H18} -- with orthogonal approaches representing the currently most relevant paradigms, edge- and vertex-centric graph processing, and evaluate our approach on their concepts.
\Cref{fig:error} shows box plots of the percentage error $e = \frac{100 \times |s - t|}{t}$ we achieve for simulation performance $s$ and ground truth performance $t$ (taken from the respective article) grouped by accelerators and algorithms.
Without single-source shortest-paths (SSSP) on HitGraph, we get a reasonable mean of $14.32\%$.
We explain why this single algorithm performs so much worse and why there are outliers for AccuGraphs weakly-connected components (WCC) algorithm in \cref{sec:reproducibility}.
In this work, we make the following contributions:
\vspace{-0.25cm}
\begin{enumerate}
    \itemsep0em
    \item We propose a \emph{simulation environment} for graph processing accelerator engineering and \emph{memory access abstractions} based on our hypothesis.
    \item We conduct a \emph{comprehensive reproducibility study} for the two representative graph processing accelerators HitGraph and AccuGraph and uncover deficiencies in performance measurement practices.
    \item We show the reduced effort of engineering new ideas with our simulation environment by example of \emph{two novel optimizations to AccuGraph}.
\end{enumerate}
\vspace{-0.25cm}

This article is structured as follows.
In \cref{sec:background} we introduce basic concepts of graph processing, FPGA-addressable DRAM, and DRAM simulation.
In \cref{sec:patterns} we conceptually specify the simulation environment, request flow abstractions, and their application to HitGraph and AccuGraph.
In \cref{sec:evaluation} we reproduce and compare the performance measurements of HitGraph and AccuGraph.
We show the engineering benefits of our approach in \cref{sec:engineering}, before discussing related work in \cref{sec:relatedwork} and concluding in \cref{sec:discussion}.

\vspace{-.3cm}
\section{Background}
\label{sec:background}
\vspace{-.3cm}
We start this section by briefly specifying graphs, how graph processing can be implemented, and what problems can be solved on graphs.
Thereafter, we shortly introduce the memory hierarchy of FPGAs and more specifically how DRAM works internally.
Lastly, we motivate the selection of Ramulator \cite{journals/cal/KimYM16} as our DRAM simulator and briefly explain how Ramulator models memory and is configured for our purpose.

\vspace{-.3cm}
\subsection{Graph Processing}
\label{sec:graph}
\vspace{-.3cm}
A graph $G = (V, E)$ is an abstract data structure consisting of a vertex set $V$ and an edge set $E \subseteq V \times V$.
Intuitively, they are used to describe a set of entities (vertices) and their relations (edges).
\begin{figure}[bt]
	\centering
	\includegraphics[width=\linewidth]{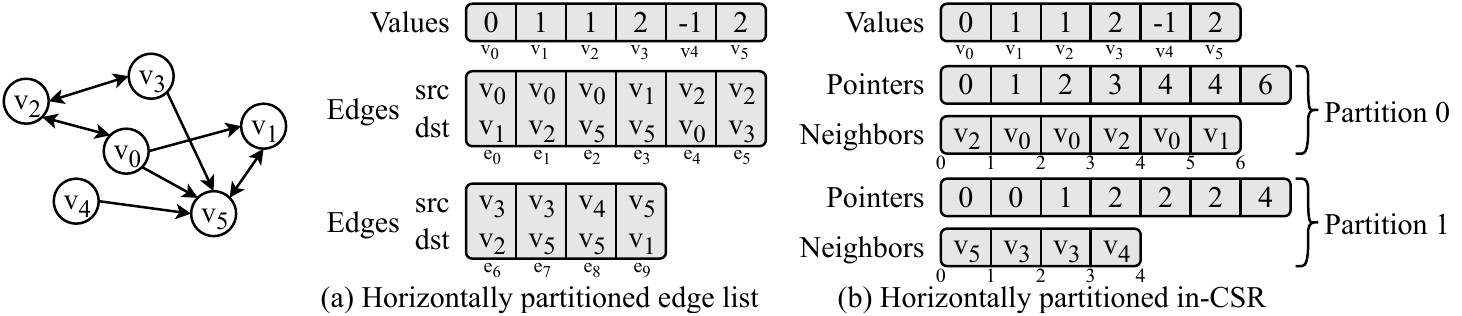}
	\caption{Graph partitioning and data structures}
	\label{fig:graph}
\end{figure}
\Cref{fig:graph} shows two possible data structure representations (both with two partitions) of the example graph.
Horizontally partitioned means dividing up the vertex set of the graph into intervals and assigning edges to the partition which interval contains their source vertex.
\Cref{fig:graph}a shows the example graph as a horizontally partitioned edge list (used by HitGraph \cite{journals/tpds/ZhouKPSW19}), which stores the graph as arrays of edges with a source and a destination vertex.
For example, edge $e_0$ connects source $v_0$ to destination vertex $v_1$.
\Cref{fig:graph}b shows the same graph as a horizontally partitioned compressed sparse row (CSR) format of the inverted edges (used by AccuGraph \cite{conf/IEEEpact/Yao0L0H18}), meaning all source and destination vertices of the edges in $E$ are swapped before building a CSR data structure on them.
CSR is a data structure for compressing sparse matrices (in this case the adjacency matrix of the graph) with two arrays. 
The values of the pointers array at position $i$ and $i + 1$ delimit the neighbors of $v_i$ stored in the neighbors array.
For example, for $v_5$ in partition $1$ the neighbors are the values of the neighbors array between $2$ and $4$, \ie $v_3$ and $v_4$.

Depending on the underlying graph data structure, graphs are processed based on two fundamentally different paradigms: edge- and vertex-centric graph processing.
Edge-centric systems (\eg HitGraph) iterate over the edges as primitives of the graph on an underlying edge list.
Vertex-centric systems iterate over the vertices and their neighbors as primitives of the graph on underlying adjacency lists (\eg CSR).
Further, for the vertex-centric paradigm, there is a distinction into push- and pull-based data flow.
A push-based data flow denotes that values are pushed along the forward direction of edges to update neighboring vertices.
A pull-based data flow (\eg applied by AccuGraph) denotes that values are pulled along the inverse direction of edges from neighboring vertices to update the current vertex.

In the context of this article, we consider the five graph problems implemented by HitGraph and AccuGraph: BFS, SSSP, WCC, sparse matrix-vector multiplication (SpMV), and PageRank (PR).
The problems specify their implementations to varying degrees.
For example, BFS denotes a sequence of visiting the vertices of a graph.
Starting with a root vertex as the frontier, in each iteration, every unvisited neighbor of the current frontier vertices is marked as visited, assigned the current iteration as its value, and added to the frontier of the next iteration.

In contrast, SSSP only specifies the desired output, \ie for each vertex $v \in V$ the shortest distance to the root vertex.
The shortest distance equals the smallest sum of edge weights of any path from the root to $v$.
If every edge is assumed to have weight 1, the result is equal to BFS.
Similarly, WCC specifies as output for each vertex its affiliation to a weakly-connected component.
Two vertices are in the same weakly-connected component if there is an undirected path between them.
There is no requirement on how these outputs are generated.

Finally, SpMV and PR specify the execution directive.
SpMV multiplies a vector (equal to $V$) with a matrix (equal to $E$) in iterations.
PR is a measure to describe the importance of vertices in a graph.
It is calculated by recursively applying $p(i) = \frac{1 - d}{|V|} + \sum_{j \in N_G(i)} \frac{p(j)}{d_G(j)}$ for each $i \in V$ with damping factor $d$, neighbors $N_G$ and degree $d_G$.

\vspace{-.3cm}
\subsection{Memory Hierarchies of Field Programmable Gate Arrays}
\label{sec:ram}
\vspace{-.3cm}
As a processor architecture platform, FPGA chips map custom architecture designs (\ie a set of logic gates and their connection) to a grid of resources (\eg look-up tables, flip-flops, and BRAM) connected with a programmable interconnection network.
The memory hierarchy of FPGAs is split up into on-chip and off-chip memory.
On-chip, FPGAs contain BRAM in the form of SRAM memory components.
On modern FPGAs, there is about as much BRAM as there is cache on modern CPUs (all cache levels combined), but contrary to the fixed cache hierarchies of CPUs, BRAM is memory finely configurable to the application.
For storage of larger data structures, DRAM (\eg DDR3\footnote{JESD79-3 DDR3 SDRAM Standard} or DDR4\footnote{JESD79-4 DDR4 SDRAM Standard}) is attached as off-chip memory.
Subsequently, we briefly introduce the internal structure of DDR3 and DDR4 to understand its implications on graph processing.

\begin{figure}[bt]
	\centering
	\includegraphics[width=0.95\linewidth]{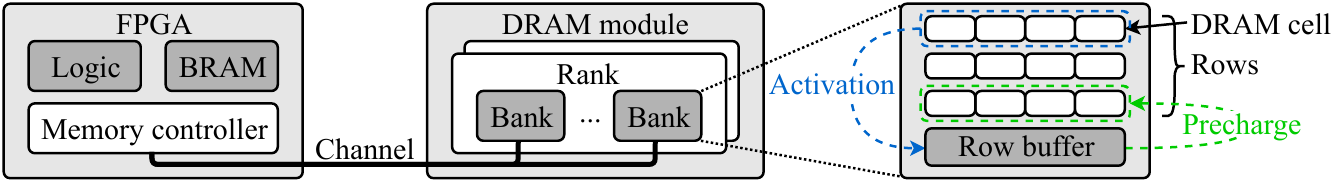}
	\caption{DRAM (adapted from \cite{journals/cal/KimYM16})}
	\label{fig:ram}
\end{figure}
The internal organization of DDR3 memory is shown in \cref{fig:ram}, which at the lowest level contains DRAM cells each representing one bit.
The smallest number of DRAM cells (\eg $16$) that is addressable is called a column.
Several thousand (\eg $1,024$) columns are grouped together into rows.
Further, independently operating banks combine several thousand (\eg $65,536$) rows with a row buffer each.

Requests to data in a bank are served by the row buffer based on three scenarios:
\begin{inparaenum}[\it (1)]
    \item When the addressed row is already buffered, the request is served with low latency (\eg $t_{CL}$: $11$ns).
    \item If the row buffer is empty, the addressed row is first activated (\eg $t_{RCD}$: $11$ns), which loads it into the row buffer, and then the request is served.
    \item However, if the row buffer currently contains a different row from a previous request, the current row has to be first pre-charged (\eg $t_{RP}$: $11$ns) and only then the addressed row can be activated and the request served.
\end{inparaenum}
Additionally, there is a minimum latency between switching rows (\eg $t_{RAS}$: $28$ns).
Thus, for high performance, row switching should be minimized.

Since one bank does not provide sufficient bandwidth, $8$ parallel banks further form a rank.
Multiple ranks operate in parallel but on the same I/O pins, thus increasing capacity of the memory, but not bandwidth.
Finally, the ranks of the memory are grouped into channels.
Each channel has its own I/O pins to the FPGA such that the bandwidth linearly increases with the number of channels.
DDR4 contains another hierarchy level called bank groups, which group two to four banks to allow for more rapid processing of commands.

\begin{figure}[bt]
	\centering
	\includegraphics[width=0.7\linewidth]{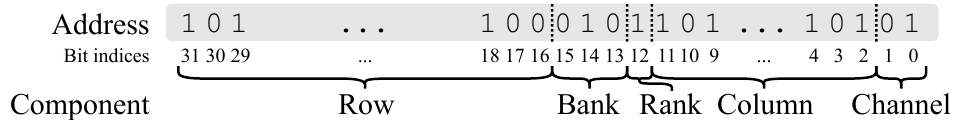}
	\caption{DRAM addressing}
	\label{fig:addressing}
\end{figure}
Data in DRAM is accessed by giving the memory a physical memory address that is split up into multiple parts internally representing addresses for each component in the DRAM hierarchy (cf. \cref{fig:addressing}).
Based on this, different addressing schemes are possible.
An example addressing scheme that aids distribution of requests over channels might first address the channels, meaning subsequent addresses go to different channels, then address columns, ranks, banks, and rows.
To further improve memory bandwidth, modern DRAM returns multiple bursts of data for each request (also called prefetching).
For DDR3 and DDR4, each request returns a total of $64$ Bytes over $8$ cycles which we call a cache line in the following.

\vspace{-.3cm}
\subsection{DRAM Simulators -- Ramulator}
\vspace{-.3cm}
To speed up the engineering of graph processing on FPGA accelerators, a DRAM Simulator is an integral part of our simulation environment (cf. \cref{fig:idea}).
For our purposes we need a DRAM simulator that supports DDR3 (for HitGraph \cite{journals/tpds/ZhouKPSW19}) and DDR4 (for AccuGraph \cite{conf/IEEEpact/Yao0L0H18}).
We chose Ramulator \cite{journals/cal/KimYM16} for this work over other alternatives like DRAMSim2 \cite{journals/cal/RosenfeldCJ11} and USIMM \cite{ChatterjeeBSPUSSAC12} because -- to the best of our knowledge -- it is the only DRAM simulator which supports (among many others like LPDDR3/4 and HBM) both of those DRAM standards (DDR3/4).

Ramulator models DRAM as a tree of state machines (\eg channel, rank, bank in DDR3) where transitions are triggered by user or internal commands.
However, Ramulator does not make any assumptions about data in memory.
Purely the request and response flow is modelled with requests flowing into Ramulator and responses being called back.
The Ramulator configuration parameters that are relevant to our work are:
\begin{inparaenum}[\it (1)] 
    \item DRAM standard,
    \item channel count,
    \item rank count,
    \item DRAM speed specification,
    \item DRAM organization.
\end{inparaenum}

\vspace{-.3cm}
\section{Memory Access Simulation Environment}
\label{sec:patterns}
\vspace{-.3cm}
In this section, we first introduce the simulation environment based on the implications of our hypothesis and show the abstractions we developed to implement memory access patterns.
Thereafter, we show how this can be applied to real graph processing accelerators.
As motivated in \cref{sec:intro}, we chose HitGraph \cite{journals/tpds/ZhouKPSW19} and AccuGraph \cite{conf/IEEEpact/Yao0L0H18} for that.

\vspace{-.3cm}
\subsection{Simulation Environment}
\label{sec:environment}
\vspace{-.3cm}
As we established in \cref{sec:intro}, one of the main challenges with evaluating new graph processing ideas on FPGAs is time-consuming and difficult development of the accelerator.
Thus, the goal of our simulation environment is reducing development time and complexity within reasonable error when compared to performance measurements on hardware.
To achieve this goal we relax the necessity of cycle accurate simulation of on-chip data flow due to our hypothesis:
\emph{Memory access patterns dominate the overall runtime of graph processing such that disregarding the internal data flow results in a reasonable error of a simulation.}
Modelling the off-chip memory access pattern means modelling request types, request addressing, request amount, and request ordering.
Request type modelling is trivial since it is obvious if requests either read or write data.
For request addressing, we assume that the different data structures (\eg edge list and vertex values) are stored adjacently in memory as plain arrays.
We generate memory addresses according to this memory layout and the width of the arrays types in Bytes.
Request amount modelling is mostly based on the size $n$ of the vertex set, the size $m$ of the edge set, average degree \emph{deg}, and partition number $p$.
We only simulate request ordering through mandatory control flow caused by data dependencies of requests.
We assume that computations and on-chip memory accesses are instantaneous by default.
In the following we introduce memory abstractions we developed for modelling request and control flow.

\begin{figure}[bt]
	\centering
	\includegraphics[width=0.8\linewidth]{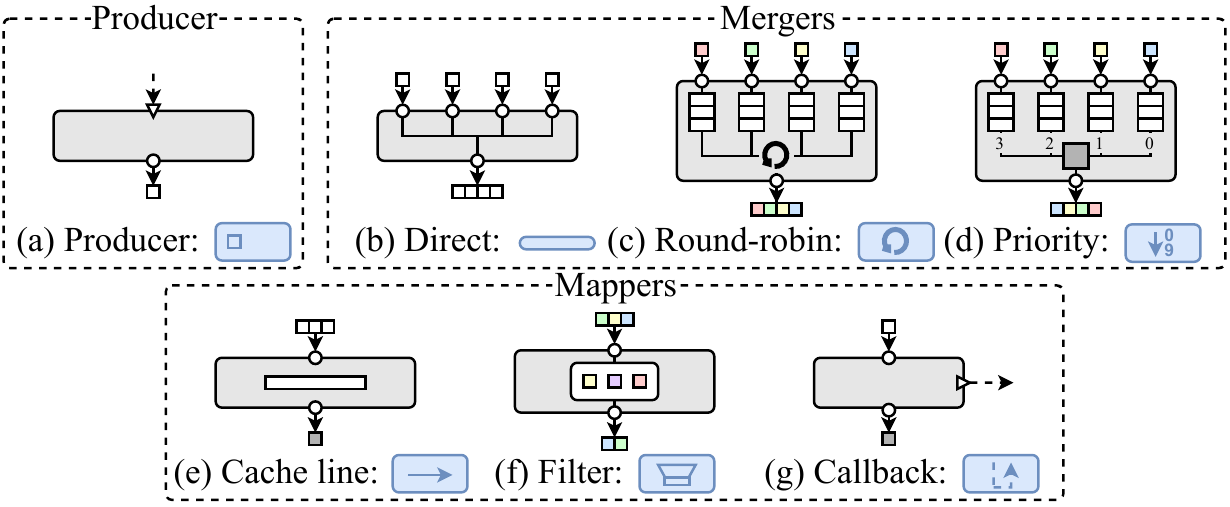}
	\caption{Memory access abstractions}
	\label{fig:patterns}
\end{figure}
\Cref{fig:patterns} shows an overview of the memory access abstractions and their icons grouped by their role during memory access as \emph{producer}, \emph{merger}, and \emph{mapper}.

\labeltitle{Producer} At the start of each request stream, a \emph{producer} (\cref{fig:patterns}a) is used to turn control flow triggers (dashed arrow) into a request stream (solid arrow).
The producer might be rate limited, but if only a single producer is working at a time or requests are load balanced down-stream, the requests are just created in bulk.

\labeltitle{Mergers} Multiple request streams might then be merged with \emph{mergers}, since Ramulator only has one endpoint.
We have deduced abstractions to merge requests in a \emph{direct} (\cref{fig:patterns}b), \emph{round-robin} (\cref{fig:patterns}c), and \emph{priority}-based (\cref{fig:patterns}d) fashion.
If there are multiple request streams that do not operate in parallel, direct merging is applied.
If request streams should be equally load-balanced, round-robin merging is applied.
If request streams should take precedence over each other, priority merging is applied.
For this, a priority is assigned to each request stream and requests are merged based on that.

\labeltitle{Mappers} Additionally to request creation with producers and ordering with mergers, we also found abstractions for request \emph{mappers}.
Thus, we introduce \emph{cache line} buffers (\cref{fig:patterns}e) for sequential or semi-sequential accesses that merge subsequent requests to the same cache line into one request.
We do buffering such that multiple concurrent streams of requests benefit from it independently by placing it as far from the memory as necessary to merge the most requests.
For data structures that are placed partially in on-chip memory (\eg prefetch buffers and caches), and thus partially not require off-chip memory requests, we introduce request \emph{filters} (\cref{fig:patterns}f) that discard unnecessary requests.
For control flow, we use a \emph{callback} (\cref{fig:patterns}g) abstraction.
We disregard any delays in control flow propagation and just directly let the memory call back into the simulation.
If requests are served from a cache line or filter abstraction, the callback is executed instantly.

In our simulation environment we instantiate a graph processing simulation and a Ramulator instance, and tick them according to their respective clock frequency.
For graph processing simulation we focus on configurability of all aspects of the simulation such that we can quickly run differently parameterized performance measurements.
Our simulation works on multiple request streams that are merged into one and fed into Ramulator.
This leads us to a immensely reduced implementation time and complexity, gives us more insight into the memory, and provides portability of ideas developed in the simulation environment. 

\vspace{-.3cm}
\subsection{HitGraph}
\vspace{-.3cm}
\begin{figure}[bt]
	\centering
	\includegraphics[width=0.95\linewidth]{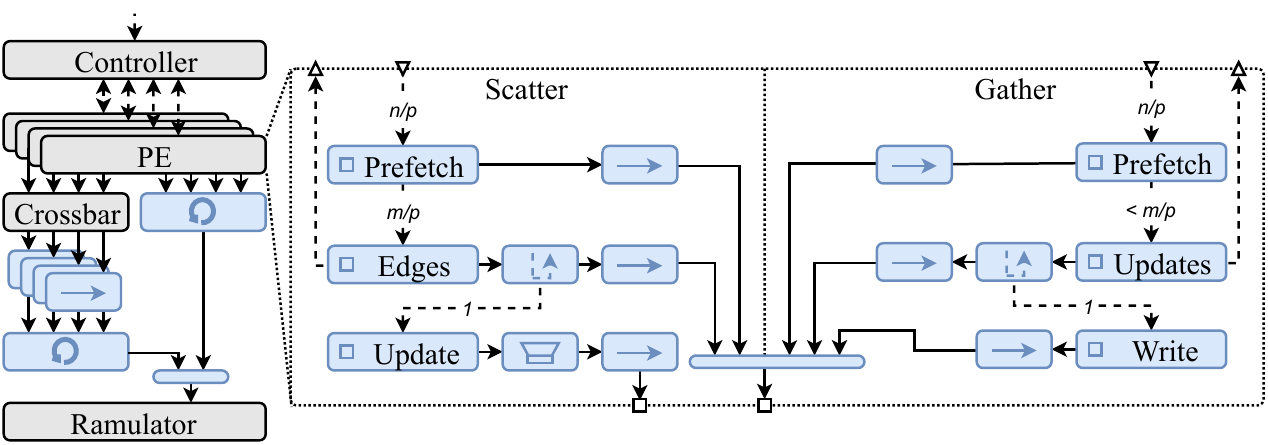}
	\caption{HitGraph request and control flow}
	\label{fig:hitgraph}
\end{figure}

HitGraph \cite{journals/tpds/ZhouKPSW19} is an edge-centric graph processing accelerator that claims to be among the best performing systems.
The basic idea is to partition the graph horizontally into $K$ partitions, stored as edge lists (cf. \cref{sec:graph}), and process the partitions in two phases in each iteration.
First, updates are produced for each edge in each partition in the scatter phase.
Second, all updates are applied to their respective vertex for each partition in the gather phase.
The main goal behind this approach is to completely eliminate random reads to data and largely reduce the amount of random writes such that only semi-random writes remain.
All reads to values of vertices are served from the prefetched partition in BRAM and all reads to either edges or updates are sequential.
Writing updates is sequential, while writing values is the only semi-random memory access.
\Cref{fig:hitgraph} shows the request and control flow modelling with our simulation environment.
Execution starts with triggering a controller that itself triggers iterations of edge-centric processing until there were no changes to vertex values in the previous iteration.
In each iteration, the controller first schedules all partitions for the scatter phase, before scheduling all partitions to the gather phase.
Partitions are assigned beforehand to channels of the memory (four channels in \cite{journals/tpds/ZhouKPSW19}) and there is a processing element (PE) for each channel.
After all partitions are finished in the gather phase, the next iteration is started or the accelerator terminates.

\labeltitle{Scatter} The scatter phase starts by prefetching the $\frac{n}{K}$ values of the current partition into BRAM.
Those requests go to a cache line abstraction, such that requests to the same cache line do not result in multiple requests to Ramulator.
After all requests are produced, the prefetch step triggers the edge reading step that reads all $\frac{m}{K}$ edges of the partition.
This is only an average value since the exact number of edges in a partition might vary because of skewed vertex degrees.
For each edge request, we attach a callback that triggers producing an update request and merge them with a cache line abstraction.
The update requests might be filtered by an optimization resulting in less than one update per edge.
The target address depends on its destination vertex that can be part of any of the $K$ partitions.
Thus, there is a crossbar that routes each update request to a cache line abstraction for each partition, which sequentially writes it into a partition-specific update queue.
After all edges have been read, the edge reader triggers the controller, which either triggers the next partition or waits on all memory requests to finish before switching phases.

\labeltitle{Gather} Similar to scatter, the gather phase starts with prefetching the $\frac{n}{K}$ vertex values sequentially.
After value requests have been produced, the prefetcher triggers the update reader, which sequentially reads the update queue written by the scatter phase before.
For each update we register a callback that triggers the value write.
The value writes are not necessarily sequential, but especially for iterations where a lot of values are written, there might be a lot of locality.
Thus, new values are passed through a cache line abstraction.

\labeltitle{Parallelization} All request streams in each PE are just merged directly into one stream without any specific merging logic, since mostly only one producer is producing requests at a time.
However, edge and update reading is rate limited to the number of pipelines in each PE (which is set to $8$ in the original article).
Since all PEs are working on independent channels and Ramulator only offers one endpoint for all channels combined, we employ a round robin merge of the PE requests in order not to starve any channel.
In addition, HitGraph applies optimizations to update generation.
As a first step, the edges are sorted by destination vertex in each partition.
This enables merging updates to the same destination vertex before writing them to memory, reducing the amount of updates $u$ from $u = m$ to $u \leq n \times K$, and providing locality to the gather phases value writing.
As a second optimization, an active bitmap with cardinality $n$ is kept in BRAM that saves for each vertex if its value was changed in the last iteration.
This enables update filtering, by filtering out updates from inactive vertices which saves a significant amount of update writes for most algorithm and data set combinations.
As a final optimization, partitions with unchanged values or no updates are skipped, which saves time spent for prefetching of values and edge/update reading for some algorithms.

\labeltitle{Configuration} HitGraph is parameterized with the number of PEs $p$, pipelines $q$, and the partition size $k$.
The number of PEs $p$ is fixed to the number of memory channels because each PE works on exactly one memory channel.
The pipeline count $q$ is limited by the bandwidth available per channel given as the cache line size divided by the edge size.
Lastly, the partition size is chosen such that $k$ vertices fit into BRAM.
HitGraph is able to use all available bandwidth due to fitting $p$ and $q$ to use all memory channels and whole cache lines of each channel per cycle.
Hence, adding more compute (\ie PEs or pipelines) would not help to solve the problem more efficiently which is in line with our hypothesis, \ie memory access dominates the performance.

\vspace{-.3cm}
\subsection{AccuGraph}
\vspace{-.3cm}
\begin{figure}[bt]
	\centering
	\includegraphics[width=.8\linewidth]{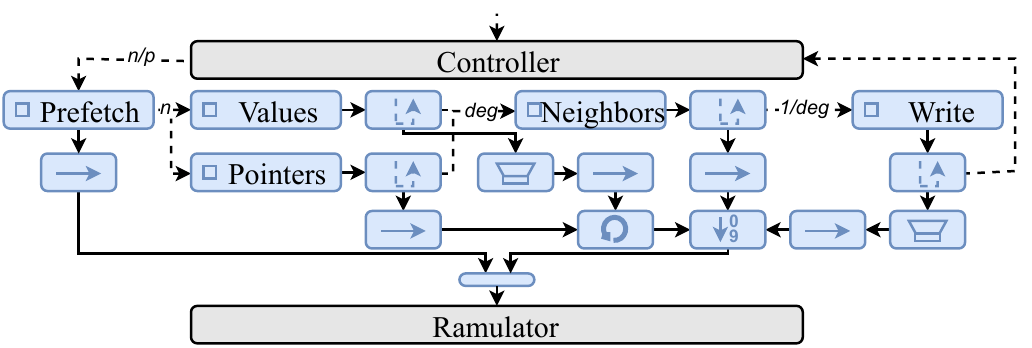}
	\caption{AccuGraph request and control flow}
	\label{fig:accugraph}
\end{figure}

AccuGraph \cite{conf/IEEEpact/Yao0L0H18} is a vertex-centric graph processing accelerator with pull data flow.
The basic idea is to partition the graph horizontally into $K$ partitions, store it as in-CSR data format, and pull updates from destination vertices (cf. \cref{sec:graph}).
The original article proposes a flexible accumulator able to merge many updates to vertex values per cycle.
\Cref{fig:accugraph} shows the request and control flow modelling of AccuGraph.
The controller is triggered to start the execution.
It iterates over the graph until there are no more changes in the previous iteration.
Each iteration triggers processing of all partitions.
Partition processing starts with prefetching the $\frac{n}{K}$ source vertex values sequentially.
Thereafter, values and pointers of all destination vertices are fetched.
The value requests are filtered by the values that are already present in BRAM from the partition prefetching.
Pointers are fetched purely sequentially.
Those two request streams are merged round robin, because a value is only useful with the associated pointers.
For every value fetched in this way, neighbors are read from memory sequentially.
Since the neighbors of subsequent vertices are in sequence in CSR, this is fully sequential.
An internal accumulator collects the changes caused through the neighbors and writes them back to memory, when all neighbors were read.
The value changes are also directly applied to the values currently present in BRAM for a coherent view of vertex values.
This is filtered such that only values that changed are written.
All of these request streams are merged by priority, with write request taking the highest priority and neighbors the second highest because otherwise the computation pipelines would be starved.
Additionally, we rate-limit neighbors loading to the number of edge pipelines present in the accelerator.

\labeltitle{Configuration} AccuGraph is parameterized by the number of vertex and edge pipelines ($8$ and $16$ in the original article) and the partition size.
Similar to HitGraph's PE and pipeline fitting, the number of edge pipelines is specifically chosen to allow processing one cache line of edges per clock cycle and thus use the entire bandwidth of the memory, again in line with our hypothesis.
The original article also describes an FPGA-internal data flow optimization which allows to approximate pipeline stalls, improving simulation accuracy significantly.
The vertex cache used for the prefetched values is partitioned into $16$ BRAM banks on the FPGA which can each serve one vertex value request per clock cycle.
Since there are $16$ edge pipelines in a standard deployment of AccuGraph, performance deteriorates quickly, when there are stalls.
Thus, we implement stalls of this vertex cache in the control flow between the neighbors and write producers.
A neighbors request callback is delayed until the BRAM bank can serve the value request.

\vspace{-.3cm}
\section{Evaluation}
\label{sec:evaluation}
\vspace{-.3cm}
In this section, we validate our simulation approach by reproducing the results reported for HitGraph \cite{journals/tpds/ZhouKPSW19} and AccuGraph \cite{conf/IEEEpact/Yao0L0H18}, by indeed showing a reasonable simulation error compared to the measurements on real FPGA hardware.
In addition, we illustrate for the first time, how these completely different graph processing approaches can be compared.

\begin{table*}[bt]
\footnotesize
\centering
\begin{tabular}{l l r r c c r r r}
 Name & Abbr. & Vertices & Edges & Dir. & Degs. & Avg. & \o & SCC \\
 \hline
 \hline
 live-journal & lj & $4,847,571$ & $68,993,773$ & \faThumbsOUp & \includegraphics[height=1em]{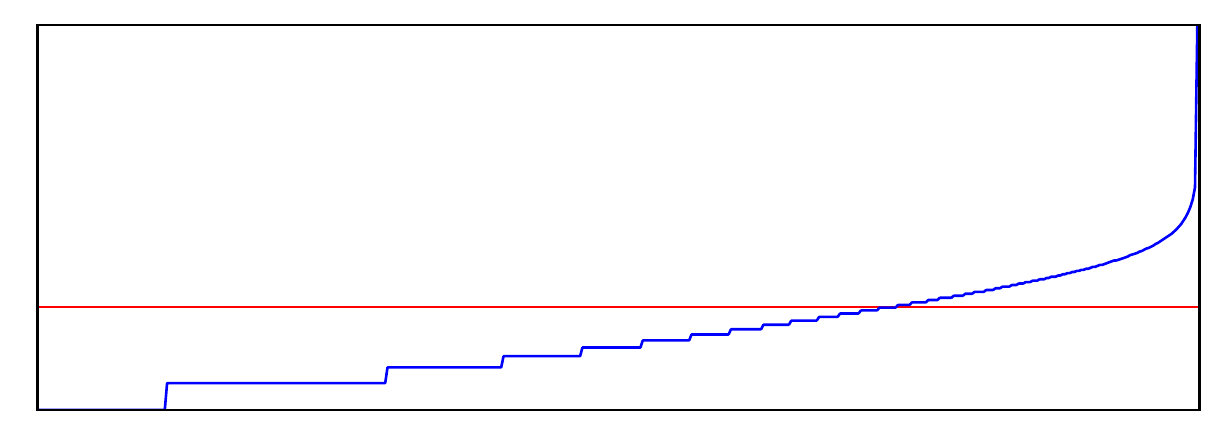} & $14.23$ & $16$ & $0.790$ \\
 wiki-talk & wt & $2,394,385$ & $5,021,410$ & \faThumbsOUp & \includegraphics[height=1em]{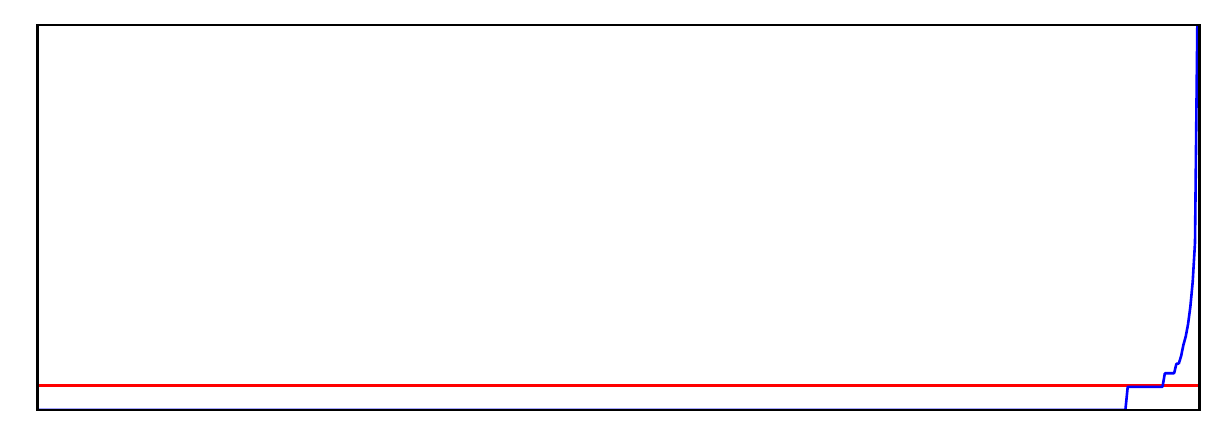} & $2.10$ & $11$ & $0.047$ \\
 \hline
 twitter\footnotemark & tw & $41,652,230$ & $1,468,364,884$ & \faThumbsOUp & \includegraphics[height=1em]{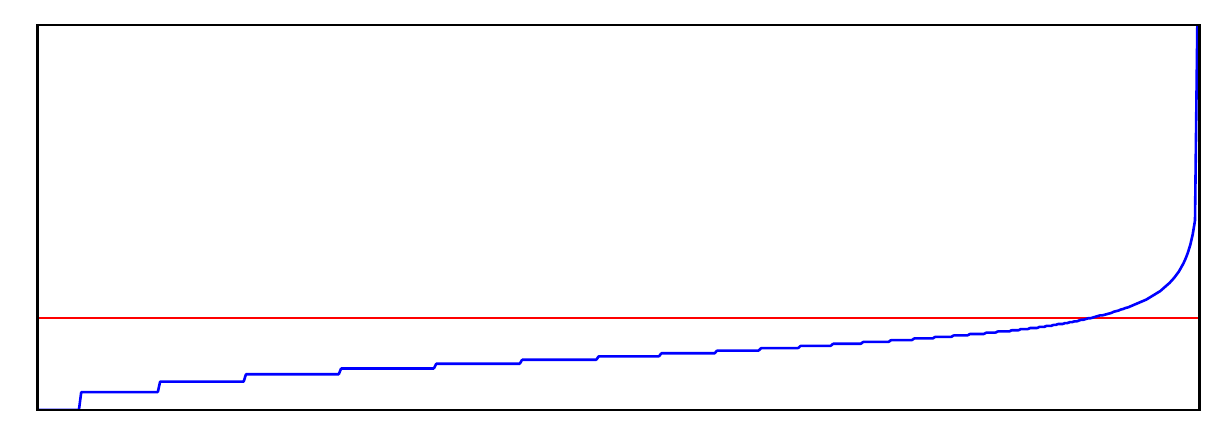} & $35.25$ & $75$ & $0.804$ \\
 rmat-24-16 & r24 & $16,777,216$ & $268,435,456$ & \faThumbsOUp & \includegraphics[height=1em]{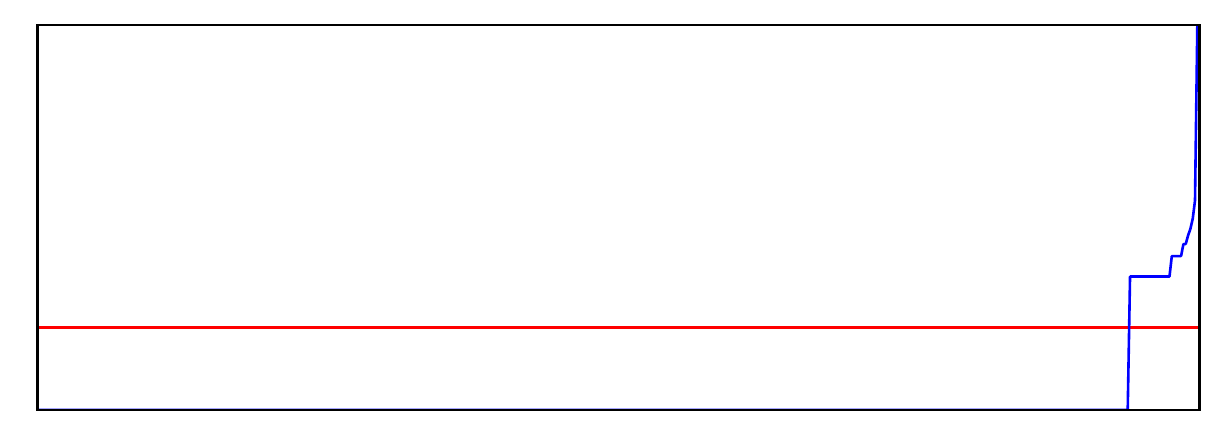} & $16.00$ & $19$ & $0.023$ \\
 rmat-21-86 & r21 & $2,097,152$ & $180,355,072$ & \faThumbsOUp & \includegraphics[height=1em]{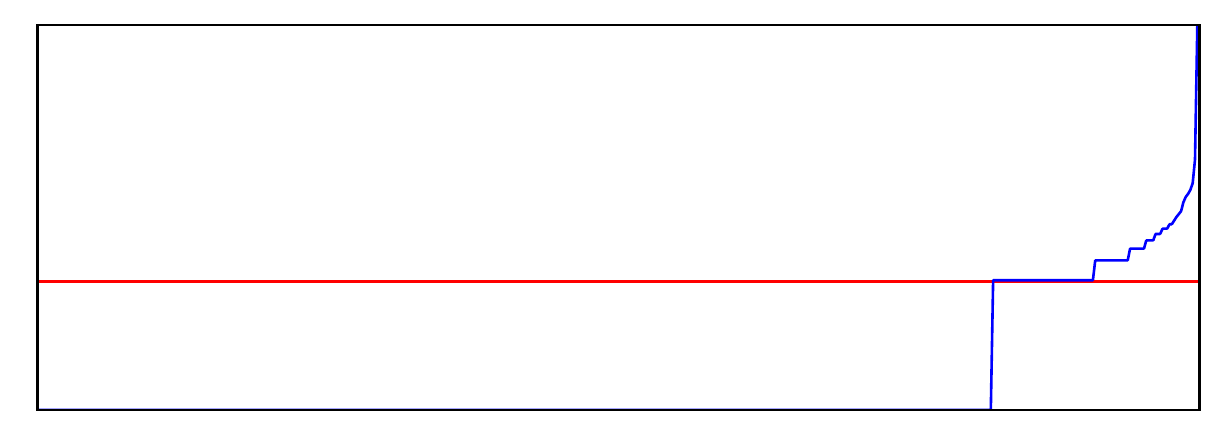} & $86.00$ & $14$ & $0.103$ \\
 roadnet-ca & rd & $1,971,281$ & $2,766,607$ & \faThumbsDown & \includegraphics[height=1em]{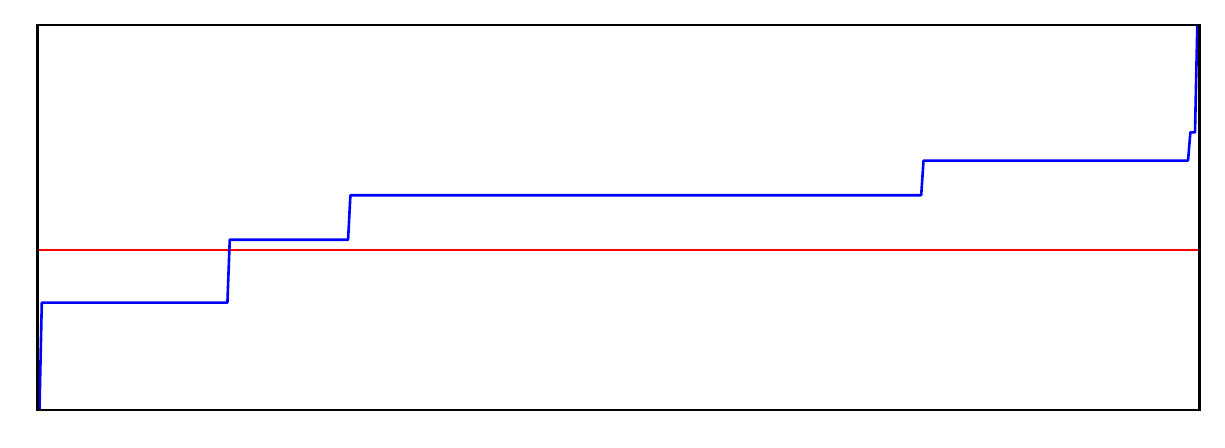} & $2.81$ & $849$ & $0.993$ \\
 berk-stan & bk & $685,231$ & $7,600,595$ & \faThumbsOUp & \includegraphics[height=1em]{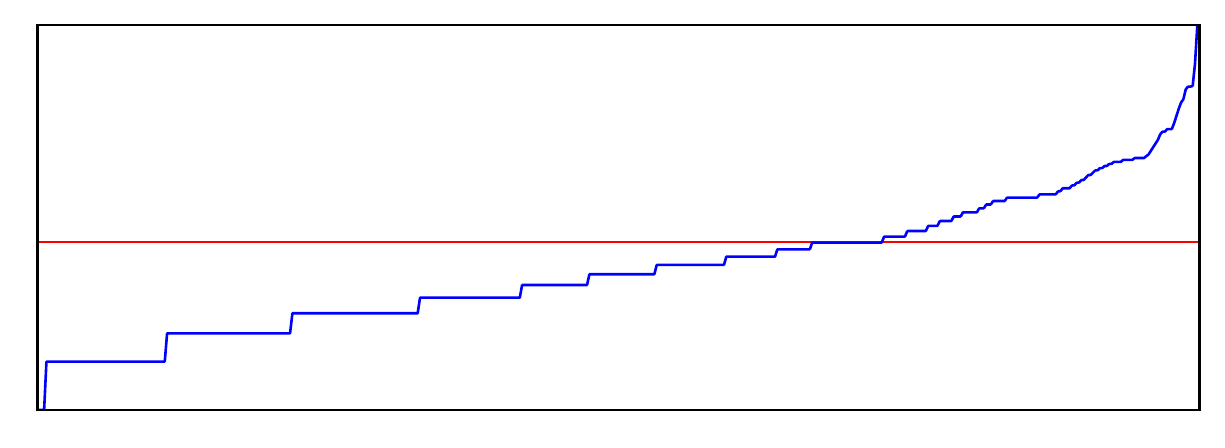} & $11.09$ & $514$ & $0.489$ \\
 \hline
 orkut & or & $3,072,627$ & $117,185,083$ & \faThumbsDown & \includegraphics[height=1em]{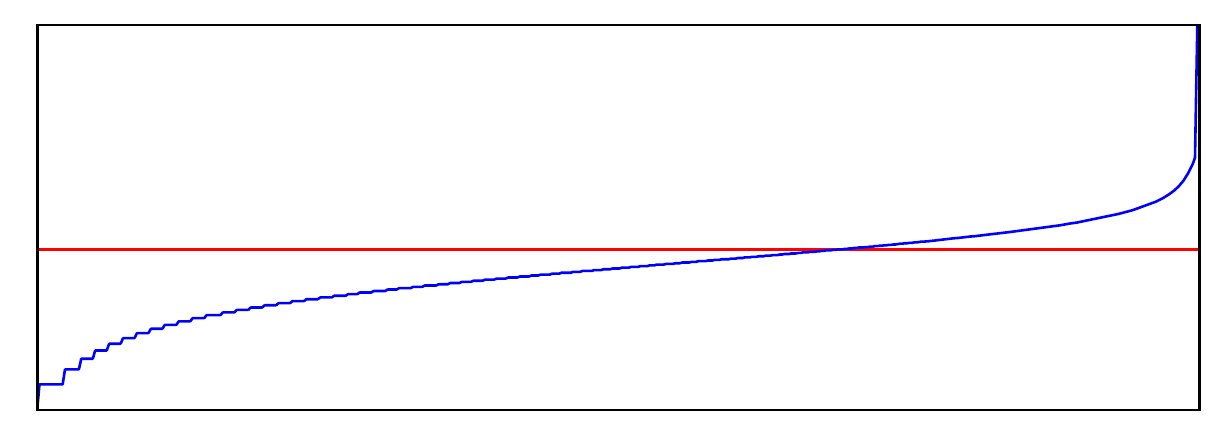} & $76.28$ & $9$ & $1.000$ \\
 youtube & yt & $1,157,828$ & $2,987,624$ & \faThumbsDown & \includegraphics[height=1em]{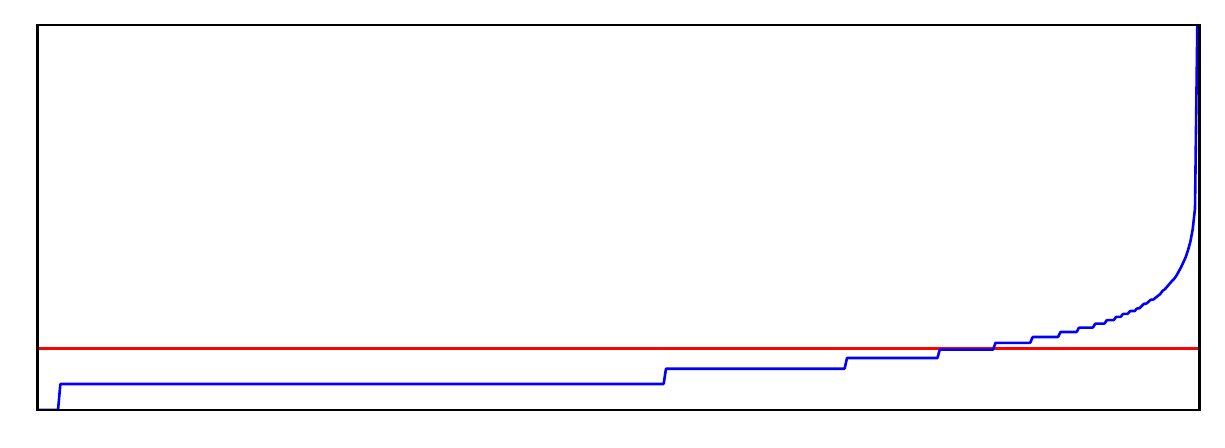} & $5.16$ & $20$ & $0.980$ \\
 dblp & db & $425,957$ & $1,049,866$ & \faThumbsDown & \includegraphics[height=1em]{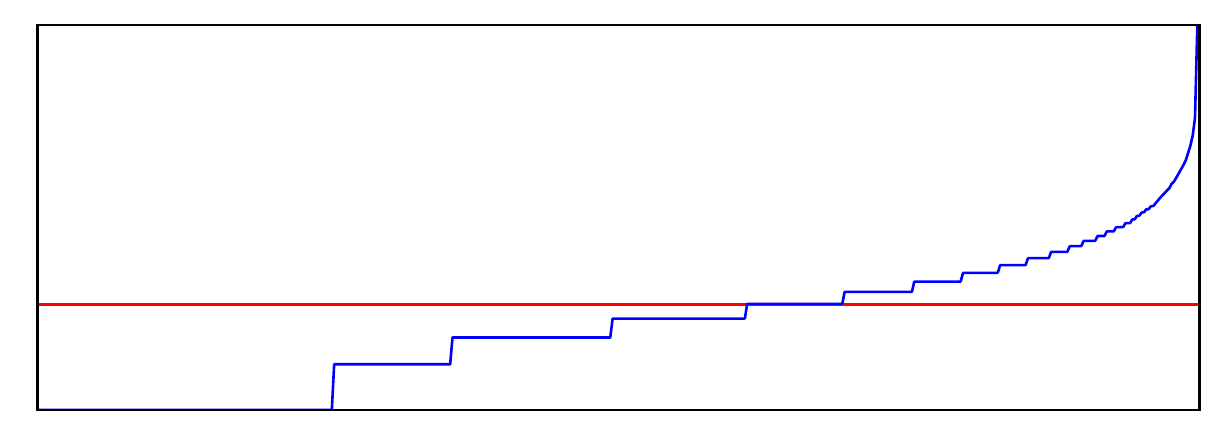} & $4.93$ & $21$ & $0.744$ \\
 slashdot & sd & $82,168$ & $948,464$ & \faThumbsOUp & \includegraphics[height=1em]{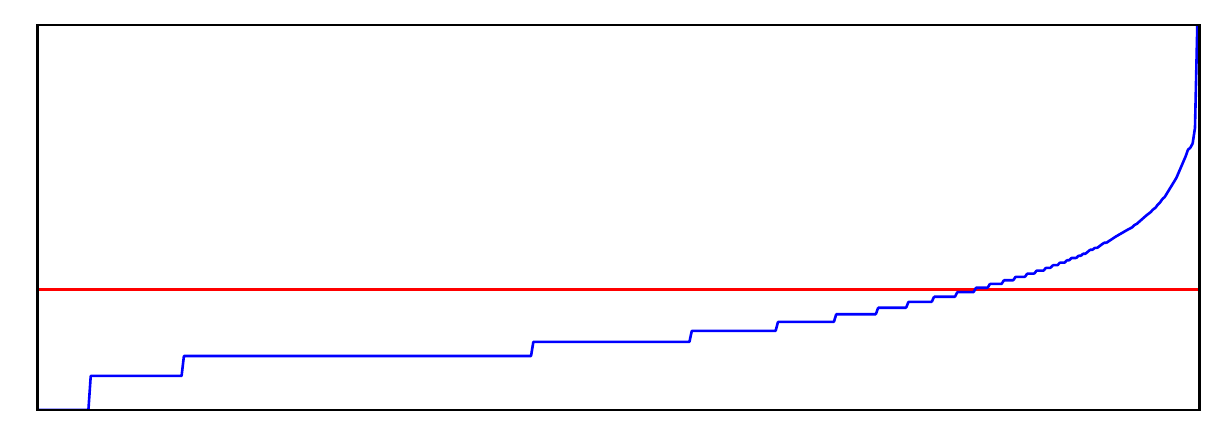} & $11.54$ & $13$ & $0.868$ \\
\end{tabular}
\medskip
\begin{tablenotes}
    \centering
    \item[*] Abbr.: Abbreviation; Dir.: Directed; Degs.: Degree distribution on log. scale; Avg.: Average degree; \o: Diameter; SCC: Ratio of vertices in the largest strongly-connected component to $n$; \faThumbsOUp: yes, \faThumbsDown: no
\end{tablenotes}
\caption{Graph data sets used by HitGraph and AccuGraph (all graphs from SNAP \cite{LeskovecK14})}
\label{tab:graphs}
\end{table*}
We take the same data sets (\Cref{tab:graphs}) and graph problems reported in the original articles to replicate their experiments.
Only the two data sets live-journal and wiki-talk are used in both articles.
HitGraph also measured performance on high diameter, constant degree graphs (\ie roadnet-ca and berk-stan) and two instances of rmat synthetic graphs.
AccuGraph measured performance on additional social graphs.
Both selections of data sets contain directed graphs, while WCC only yields correct results for undirected graphs.
This does not concern our reproducibility measurements but needs to be considered in the future.

\vspace{-.3cm}
\subsection{Reproducibility}
\label{sec:reproducibility}
\vspace{-.3cm}
\footnotetext{Not officially listed on the SNAP \cite{LeskovecK14} website anymore}
\begin{table}[bt]
\footnotesize
\centering
\begin{tabular}{l|l r r l l}
    Approach & Type & Channels & Ranks & Speed & Organization \\
    \hline
    \hline
    HitGraph & DDR3 & $4$ & $2$ & 1600K & 8Gb\_x16 \\
    AccuGraph & DDR4 & $1$ & $1$ & 2400R & 4Gb\_x16 \\
    \hline
    Comparability & DDR4 & $1$ & $1$ & 2400R & 8Gb\_x16 \\
\end{tabular}
\caption{DRAM configurations}
\label{tab:dram}
\end{table}
\begin{table}[bt]
\footnotesize
\centering
\begin{tabular}{l|c|r r r r r r r}
    Approach & Weighted & SpMV & SSSP & PR & WCC & BFS & Vertex & Pointer \\
    \hline
    \hline
    HitGraph & \faThumbsOUp & $32$ & $32$ & $32$ & $32$ & - & $32$ & - \\
    AccuGraph & \faThumbsDown & - & - & $32$ & $32$ & $8$ & $32$ & $32$ \\
    \hline
    Comparability & \faThumbsDown & - & $32$ & $32$ & $32$ & $32$ & $32$ & $32$ \\
\end{tabular}
\caption{Data structure configurations (type width in bits)}
\label{tab:data-structure}
\end{table}
\begin{table}[bt]
\footnotesize
\centering
\begin{tabular}{l|r r|r|r r r r}
    Approach & PEs & Pipelines & Elements & Vertex pipelines & Edge pipelines & VS & ES \\
    \hline
    \hline
    HitGraph & $4$ & $8$ & $256,000$ & - & - & - & - \\
    AccuGraph & - & - & $\infty$ & $8$ & $16$ & $8$ & $8$ \\
    \hline
    Comparability & $1$ & $16$ & $1,024,000$ & $8$ & $16$ & $8$ & $8$ \\
\end{tabular}
\medskip
\begin{tablenotes}
    \centering
    \item[*] PEs: Processing elements; VS: Vertex pipeline size; ES: Edge pipeline size
\end{tablenotes}
\caption{Parameter configurations}
\label{tab:parameters}
\end{table}

We measure the quality of the simulation as the percentage error $e = \frac{100 \times |s - t|}{t}$ of the simulation performance measurement $s$ compared against the ground truth $t$ reported by the respective article.
The HitGraph numbers are extracted from a table and the AccuGraph numbers are taken from a chart.
To reproduce the experiments as closely as possible, we parameterized the simulation environment according to configurations from the original articles.
\Cref{tab:dram} shows the memory configurations of the reproducibility studies and the comparability study.
\Cref{tab:data-structure} shows the data structure configurations.
HitGraph uses weighted graphs and uniformly wide value types for all problems.
AccuGraph uses unweighted graphs and an optimized 8bit-wide unsigned integer for BFS (problematic for constant degree graphs).
\Cref{tab:parameters} shows the respective graph processing accelerator parameters described in \cref{sec:environment} and how they are configured.
Both approaches share the partition size as a parameter.
For the reproducibility study, AccuGraph is assumed to fit all vertices in BRAM for BFS and only for PR and WCC measurements on live-journal and orkut, the partition size is set to $1,700,000$ vertices.

\begin{figure}[bt]
    \centering
	\includegraphics[width=0.8\textwidth]{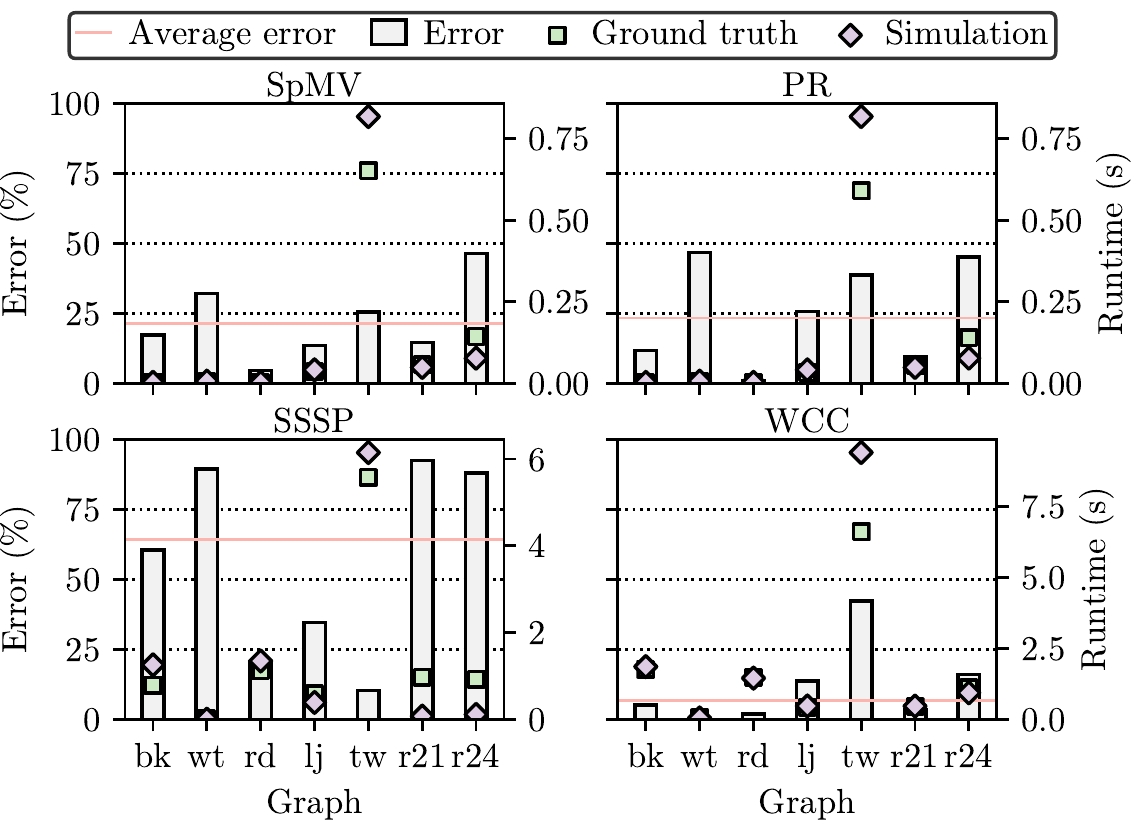}
    \caption{HitGraph measurements}
    \label{fig:hitgraph-plot}
\end{figure}
\Cref{fig:hitgraph-plot} shows the HitGraph performance measurements for SpMV, PR, SSSP, and WCC as runtime in seconds (raw numbers can be found in \cref{tab:hitgraph}).
Overall, we observe a consistent outlier in the twitter graph.
However, we notice that the HitGraph article reports the diameter of the twitter graph as being $15$, while we report it as being $75$ (cf. \cref{tab:graphs}).
Thus, we assume that our version of the graph is different and exclude it from all error averages in this article while still showing it in the plots for completeness (error source \circled{1}).
SpMV and PR result in the same simulation performance, but since ground truth values are slightly different, we get a different error.
In the original article, the authors measure only a single iteration of SpMV and PR.
However, we found that for very short runtimes of single iteration executions differences of a few cycles can already cause large deviations leading to the errors we observe for SpMV and PR (error source \circled{2}).
We advise using multiple iterations of such algorithms in benchmarks in the future.

\begin{figure}[bt]
    \centering
	\includegraphics[width=.65\textwidth]{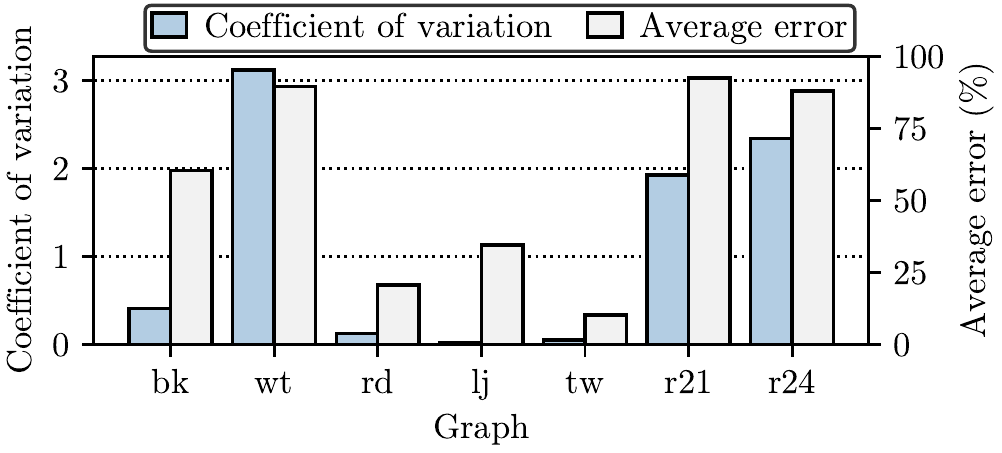}
    \caption{HitGraph SSSP runtime variation study}
    \label{fig:variance}
\end{figure}
SSSP shows by far the worst error, with some executions running much shorter in simulation than in the ground truth measurements.
This can be explained by the problem's dependence on the input root vertex (error source \circled{3}).
The HitGraph authors randomly choose $20$ root vertices and report the average runtime.
However, wiki-talk and the rmat graphs have many strongly-connected components (SSCs) with just one or a few vertices (cf. \cref{tab:graphs}).
This causes algorithms like SSSP or BFS to immediately terminate after one iteration over the graph with very little runtime which results in large variation in performance measurements for root vertices from many small and few big SSCs shown in \cref{fig:variance}.
The error is strongly correlated to the coefficient of variation in the runtimes (given by $\frac{\sigma}{\mu}$ with the standard deviation $\sigma$ and the mean $\mu$).
This leads us to the conclusion that $20$ random root vertices are not enough to stabilize the runtime measurements for graphs with such structure.
We advocate for sharing how roots are picked in the future.\footnote{We generated the $20$ random root vertices with the \texttt{mt19937} generator in C++ with seed $3483584297$.}
Moreover, the HitGraph article does not specify how edge weights are set in the graph, which can also influence runtimes of SSSP (error source \circled{4}).
We initialized all weights to $1$.
We regard WCC as the most reliable indicator for simulation quality because it does not depend on input variables and runs long enough so fixed overheads are irrelevant.
We observe a low simulation error for WCC, which reassures us that the off-chip memory access modelling works well for HitGraph.
Besides the twitter graph (which we explicitly excluded), the simulation almost perfectly matches the ground truth.

\begin{figure}[bt]
    \centering
	\includegraphics[width=\textwidth]{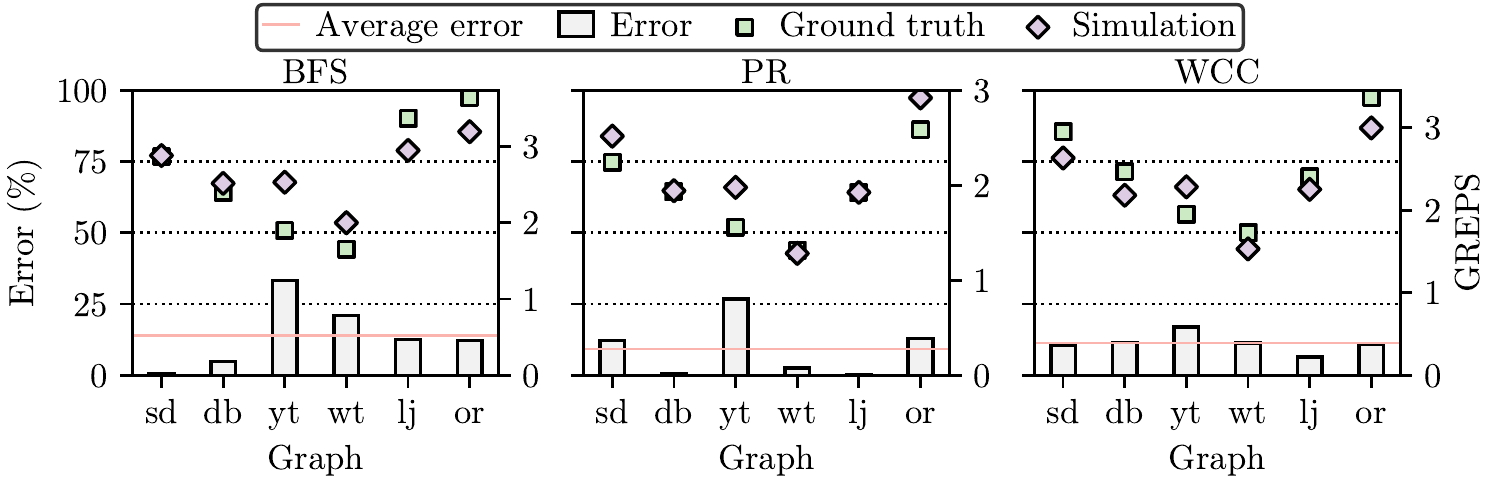}
    \caption{AccuGraph measurements}
    \label{fig:accugraph-plot}
\end{figure}
\Cref{fig:accugraph-plot} shows performance measurements for AccuGraph for BFS, PR, and WCC as billions of read edges per second (GREPS) (raw numbers can be found in \cref{tab:accugraph}).
We calculate REPS as the number of actually read edges $m \times i$ (where $i$ is the number of iterations) divided by the runtime $r$, which the original article calls traversed edges per second (TEPS).
However, this is misleading since the well-known Graph500 benchmark defines TEPS as the number of edges in a graph $m$ divided by the runtime $r$.
Thus, we rename the performance measure to REPS.
As we already saw in \cref{fig:error}, the average error is very similar for all problems and fits the relative performance of the graph data sets well.

\begin{figure}[bt]
    \centering
	\includegraphics[width=.55\textwidth]{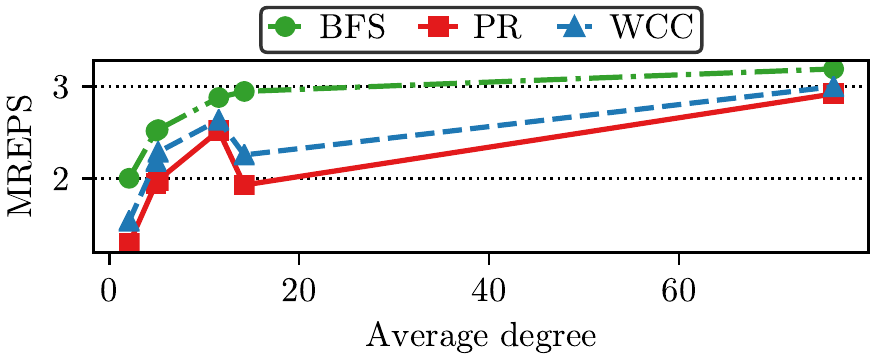}
    \caption{AccuGraph performance by average degree}
    \label{fig:average-degree}
\end{figure}
The only consistent outlier is the youtube graph which relatively performs better in all simulation measurements than is suggested by the ground truth measurements (error source \circled{5}).
The original article notes that the performance of AccuGraph logarithmically depends on the average degree of vertices which we also reproduced (cf. \cref{fig:average-degree}). 
Thus, youtube should perform the way our measurements suggest, because it has a slightly higher average degree than the dblp graph.
This may be an anomaly in the measurements performed by the AccuGraph authors.
WCC is slightly slower in our simulations than they are on the accelerator and PR is slightly faster.
There may be a fixed overhead that we are measuring in our experiments and is not measured in theirs.
The better performance of PR, however, is expected, since we do not take the longer latencies and incurred pipeline stalls of floating point arithmetics into account (error source \circled{6}).

\vspace{-.3cm}
\subsection{Comparability}
\label{sec:comparability}
\vspace{-.3cm}
\begin{figure}[bt]
	\centering
	\includegraphics[width=\linewidth]{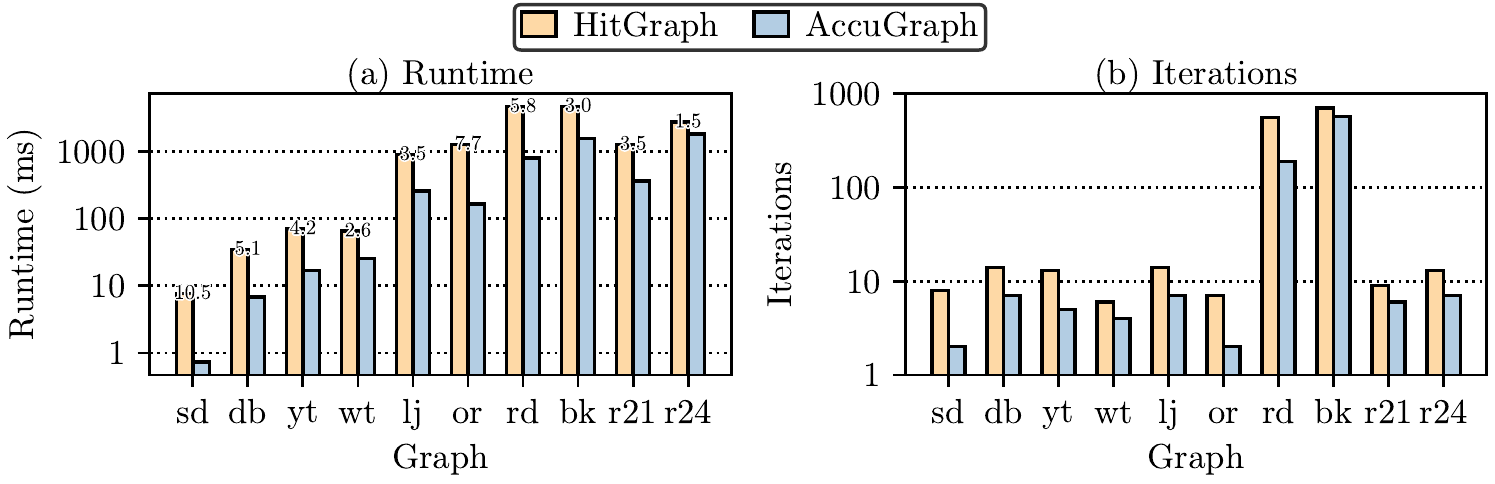}
	\caption{HitGraph vs. AccuGraph on comparable configurations (with improvement of AccuGraph over HitGraph for runtime)}
	\label{fig:comparability}
\end{figure}

With these encouraging reproducibility errors and the deeper insight in the approaches configurations, \cref{fig:comparability} shows a comparison of HitGraph and AccuGraph on an equal configuration (cf. Comparability in \cref{tab:dram} -- \cref{tab:parameters}).
It is not easily possible to use AccuGraph with weighted edges, such that we chose unweighted edges for these measurements.
Also it was not possible to expand AccuGraph to use four memory channels, such that we took the AccuGraph DRAM configuration, but increased the memory size to $8$GB to be able to accommodate the rmat graphs.
However, even this DRAM configuration does not fit the twitter graph which we thus excluded.
Moreover, we configured HitGraph to process up to $16$ edges each cycle just like AccuGraph and set the partition size to a reasonable $1,024,000$ vertices.
We show performance numbers of WCC on all graphs used in either of the original articles, since we got the lowest error for WCC.
This includes high diameter graphs (\eg roadnet and berkley-stanford) that AccuGraph has not been tested on yet.

For the two graphs that both systems were originally tested on, AccuGraph ($\sim 1728$ MREPS) reported slightly higher numbers than HitGraph ($1665$ MREPS) on wiki-talk and HitGraph ($3322$ MREPS) reported much higher numbers on live-journal than AccuGraph ($\sim 2406$ MREPS) in the original articles.
However, this is contrary to the absolute numbers we report here as runtime in seconds (\cref{fig:comparability}a).
HitGraph performs worse on all graphs (the numbers in the runtime chart are the factor calculated by dividing the HitGraph runtime by the AccuGraph runtime).
Even the simulation inaccuracy of a mean percentage error of $8.997\%$ for WCC measured in \cref{sec:reproducibility} cannot account for this.
This leads us to a first observation that REPS (used as a performance indicator in the original articles) is not a reliable performance measure due to it hiding differences in runtime.

When comparing the two approaches, we notice that AccuGraph needs fewer iterations for WCC than HitGraph (cf. \cref{fig:comparability}b). 
AccuGraph converges on a solution quicker because it updates values directly.
Due to the two-staged approach of HitGraph, it always works on the values of the past iteration.
Lower iteration count is exhibited especially by measurements on high average degree, low-diameter graphs (\eg slash-dot and orkut).
Additionally, AccuGraph shows relatively higher performance for small graphs (\eg slash-dot and dblp).
In this scenario, all vertex value reads besides the partition prefetch are served from low-latency, on-chip BRAM, because there is only one partition.
The last two factors for AccuGraphs higher performance are: HitGraph needs more requests to read the edges of the graph \emph{and} the updates, and HitGraph reads $64$ bits per edge while AccuGraph only reads $> 32$ bits per edge due to the CSR format.
We thus expect a performance advantage of at least factor $2$ on all measurements which is not achieved by AccuGraph on the rmat-24-16 graph.
This is due to the partition skipping optimization of HitGraph (not available for AccuGraph).
This leads to our second observation that AccuGraph has a categorical advantage over HitGraph because of its direct application of value changes and compressed graph format.

\vspace{-.3cm}
\subsection{Summary -- Error Analysis}
\vspace{-.3cm}

We saw that our simulation environment is able to reproduce the ground truth performance measurements of the original articles with reasonable error (\cref{sec:reproducibility}).
This is possible for bandwidth-bound accelerators (like HitGraph and AccuGraph) despite the radical hypothesis of disregarding FPGA internals.
Especially if relative performance behaviour of approaches is so significantly different (cf. \cref{sec:comparability}), an average error of \eg $8.997\%$ for WCC is reasonable to make sound relative comparisons.
However, we also identified six sources of errors which we discuss in the following.
For measurements with insufficiently specified input parameters like start vertices (error source \circled{3}) and edge weights (error source \circled{4}) we see large errors for some graphs.
Additionally, we attribute at least some of the error to noise in the measurements.
For example, very low runtime measurements like individual iterations of SpMV and PR (error source \circled{2}) can lead to significant noise.
We see underestimation of runtime due to missing modelling of pipeline bubbles that slow down request generation or missing modelling of \eg floating point units that perform complicated calculations (error source \circled{6}).
Lastly, there remain two graphs in twitter and youtube for which we cannot explain performance differences based on our simulation but rather attribute these differences to different data sets or different usage of them (error sources \circled{1} and \circled{5}).

One not easily quantifiable, possible error source (error source \circled{7}) we want to add here is interpretation based on understanding of the original article's description of their approach.
This was \eg especially necessary for data structures with missing data type specifications.
To aid researchers trying to understand the approaches we specified the data types in \cref{tab:data-structure} and advise to completely specify such parameters in the future to aid reproduction of results.

Without SSSP, we see a low mean error of $14.32\%$.
Thus, for certain use cases, we confirm our hypothesis: \emph{Memory access patterns dominate the overall runtime of graph processing such that disregarding the internal data flow results in a reasonable error of a simulation.}
We advise that the simulation should be used in use cases where relative performance behaviour is compared rather than where absolute performance should be estimated.
Additionally, if the relative performance behaviour is close for the compared approaches our simulation approach might lead to inaccurate conclusions.

\vspace{-.3cm}
\section{Example for Faster Graph Accelerator Engineering}
\label{sec:engineering}
\vspace{-.3cm}
\begin{figure}[bt]
	\centering
	\includegraphics[width=\linewidth]{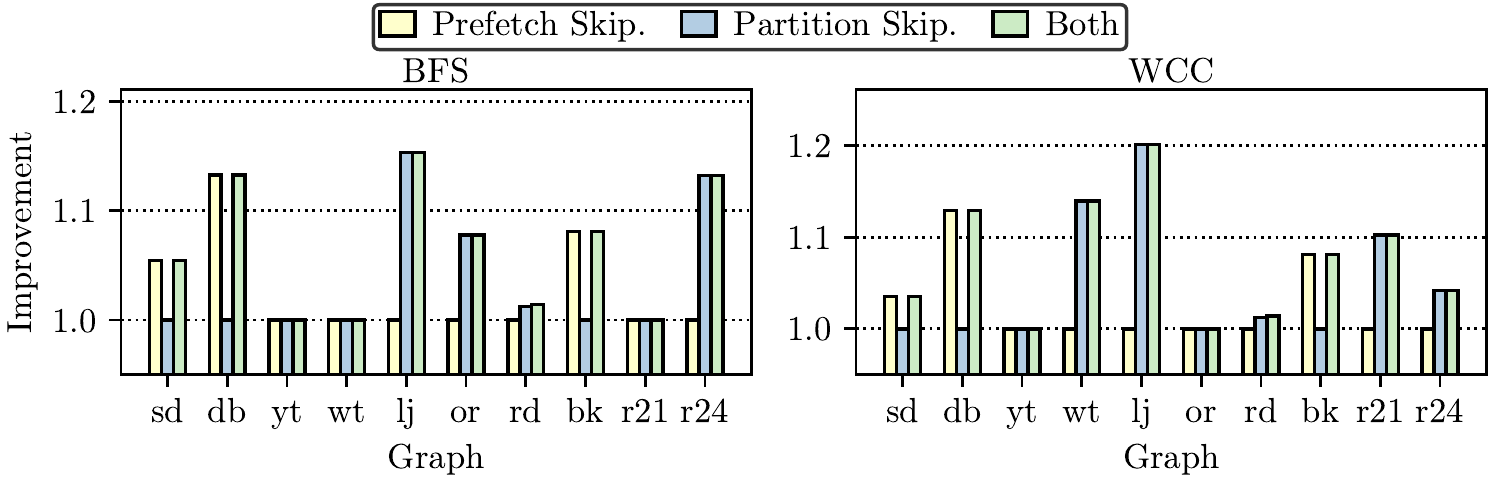}
    \caption{Runtime improvement of optimizations over baseline}
    \label{fig:engineering}
\end{figure}
In this section, we illustrate how our approach helps to speed up graph processing accelerator engineering by the example of two enhancements of AccuGraph that we found while analyzing the performance in the previous section.
Note that instead of implementing the enhancements on the FPGA itself, our simulation approach is used to quickly assess the altered designs for the different data sets as well as potentially different DRAM types, thus reducing the overall engineering time by a form of rapid graph accelerator prototyping.

\labeltitle{Enhancement ideas} AccuGraph writes all value changes through to off-chip memory and also applies them to BRAM if they are in the current partition.
Thus, BRAM and off-chip memory are always in sync.
Nevertheless, at the beginning of processing a partition, the value set is prefetched even if the values are already present in BRAM.
Thus, the first optimization we propose is prefetch skipping in this case.
Especially for the rmat-24-16 graph we also saw the effectiveness of partition skipping with HitGraph (cf. \cref{fig:comparability}).
Thus as a second optimization, we propose adding partition skipping to AccuGraph.
Both optimizations can easily be added to AccuGraphs control flow by directly triggering the value and pointer reading producers or completely skipping triggering of execution for certain partitions respectively.
For prefetch skipping we compare the currently fetched partition with the next partition to prefetch and skip prefetching if they are the same.
For partition skipping we keep track if any value of the vertices of a partition were changed and skip the partition if none changed.
The optimizations also work in combination.

\labeltitle{Results} To prove their effectiveness, we measure the effect of both optimizations for BFS and WCC separately and combined (\cref{fig:engineering}).
For all small graphs with only one partition we see an improvement based on prefetch skipping.
Partition skipping is not applicable to those graphs.
For some other graphs we see an improvement based on partition skipping.
Prefetch skipping only sometimes contributes a small improvement but only when combined with partition skipping.
PR as a stationary algorithm is not shown, since no partitions can be skipped by definition.
For prefetch skipping there are similar performance improvements on PR compared to BFS and WCC.
Overall we see no decrease in performance, suggesting that both optimizations should always be applied.

Note that these insights on the two enhancement ideas were possible in a relatively short amount of time, compared to engineering on an actual FPGA.
Developing and verifying a complicated FPGA design usually takes weeks, while the implementation of a new graph accelerator approach in our simulation environment takes days or even just hours if the approach is well understood before.
Additionally, the iteration time is much improved.
Synthesis runs for compiling hardware description code to FPGA take hours up to a day without many possibilities of incremental synthesis, while a complete compilation of our simulation environment takes $33.5$ seconds on a server with the possibility of easily utilizing parameters and incremental compilation.
As a downside, the simulation runs longer than a synthesized design on an FPGA would.
However, the user is not limited by special hardware that is only available in limited numbers (FPGAs).
Many runs can be executed in parallel on one or even multiple servers.
Especially for the very fragmented FPGA market, virtualized offers for FPGAs might not be available for specific boards.

\vspace{-.3cm}
\section{Related Work}
\label{sec:relatedwork}
\vspace{-.3cm}
To the best of our knowledge, there are no prior works on only using the off-chip memory requests paired with a DRAM simulator to make graph processing accelerators more comprehensible and performance measurements reproducible and comparable.

\labeltitle{Cache miss runtime estimation}
\cite{conf/vldb/ManegoldBK02} describe a cost model to approximate query runtimes in relational databases based on cache misses of memory requests.
They focus on CPU cache hierarchies which allow much less fine-granular data placement than FPGA memory hierarchies (cf. \cref{sec:ram}).
Additionally, they do not perform simulations of requests but model performance theoretically based on the model parameters of number of cache misses and cache latency not applicable to FPGAs.

\labeltitle{Comprehensibility}
\cite{conf/reconfig/ZhouCP15} introduces a DRAM model and simulation for HitGraph.
The simulation also generates the sequence of requests, but instead of simulating DRAM runtime, it assumes that every request results in a row buffer hit and models the performance along the cycles needed for processing the data and approximated pipelines stalls.
However, they do not show performance numbers generated with this simulation.
\cite{conf/islped/YanHLALMDYZF019} uses Ramulator as the underlying DRAM simulator for a custom cycle-accurate simulation of the accelerator Graphicionado \cite{conf/micro/HamWSSM16}.
However, this incurs very high implementation time.

\labeltitle{Reproducibility}
Regarding reproducibility, there are prior works on ways to report performance results in such a way that it suits the own approach on parallel computing systems \cite{Davison95, conf/sc/HoeflerB15}.
The graph processing accelerator domain seems to suffer from similar problems and lack of widely accepted standards in benchmarking.

\labeltitle{Comparability}
Ramulator \cite{journals/cal/KimYM16} was previously used in a work studying the interactions of complex workloads and DRAM types \cite{journals/pomacs/GhoseLHCM19}.
They uncovered how the internal structure and characteristics of DRAM (DDR3 and DDR4 in our work) relate to performance gains or losses on otherwise fix benchmarks.
This may be a future angle to improve graph processing accelerator performance by fitting the DRAM type to the algorithms and data sets.
Similarly to our work, \cite{conf/icws/XuZLSGWLZ17} raises awareness for lacking comparability in graph processing approaches on CPU-based cloud platforms.
They find tradeoffs in approaches between different workloads and differently structured graphs.

\vspace{-.3cm}
\section{Discussion and Outlook}
\label{sec:discussion}
\vspace{-.3cm}
In this article, we propose a simulation environment for graph processing accelerator approaches based on our hypothesis:
\emph{Memory access patterns dominate the overall runtime of graph processing such that disregarding the internal data flow results in a reasonable error of a simulation.}
The simulation environment models request flow fed into a DRAM simulator (\ie Ramulator \cite{journals/cal/KimYM16}) and control flow based on data dependencies.
We developed a set of memory access abstractions and applied these to FPGA implementations (\ie HitGraph \cite{journals/tpds/ZhouKPSW19} and AccuGraph \cite{conf/IEEEpact/Yao0L0H18}) representing the two dominating graph processing approaches (\ie edge- and vertex-centric).

Even though the simulation environment disregards large parts of the graph processing accelerator, we showed that it is able to reproduce ground truth measurements with a reasonable error for most workloads.
In our analysis of the large errors for some workloads we found insufficiencies in benchmark setups and attribute some error to the radical hypothesis of our approach.
We further utilized the simulation environment to compare the two approaches on a fixed configuration, revealing insufficiencies in existing performance measurements of graph processing accelerators.
Lastly, we show that our simulation approach significantly reduces the iteration time to develop and test graph processing approaches for hardware accelerators by example of two optimizations for AccuGraph that we propose.
In addition, our approach allows for deeper inspection with DRAM statistics as well as easy parameter variation without a fixed hardware platform.

In future work, we will extend the approach to an analytical performance model and study the relationship between DRAM types (\eg HBM, HMC, or LPDDR) and workload types, as well as further graph processing accelerator approaches in more detail.
Additionally, there are open questions on how to reduce the relative errors of the simulation environment.
This could, \eg be achieved by studying the simulation environment in more depth on a graph accelerator we implemented and fully control the benchmark setup for.

\vspace{-.3cm}
\bibliography{paper}

\begin{thebibliography}{KYM16}

\bibitem[Ab19]{journals/sigmod/AbadiAABBBBCCDD19}
Abadi, Daniel; Ailamaki, Anastasia; Andersen, David; Bailis, Peter; Balazinska,
  Magdalena et~al.: The Seattle Report on Database Research.
\newblock {SIGMOD} Rec., 48(4):44--53, 2019.

\bibitem[Be19]{journals/corr/abs-1910-09017}
Besta, Maciej; Peter, Emanuel; Gerstenberger, Robert; Fischer, Marc;
  Podstawski, Michal; Barthels, Claude et~al.: Demystifying Graph Databases:
  Analysis and Taxonomy of Data Organization, System Designs, and Graph
  Queries.
\newblock CoRR, abs/1910.09017, 2019.

\bibitem[BRS13]{journals/queue/BaconRS13}
Bacon, David~F.; Rabbah, Rodric~M.; Shukla, Sunil: {FPGA} Programming for the
  Masses.
\newblock {ACM} Queue, 11(2):40, 2013.

\bibitem[Ch12]{ChatterjeeBSPUSSAC12}
Chatterjee, Niladrish; Balasubramonian, Rajeev; Shevgoor, Manjunath; Pugsley,
  Seth; Udipi, Aniruddha; Shafiee, Ali; Sudan, Kshitij; Awasthi, Manu; Chishti,
  Zeshan: USIMM: the Utah SImulated Memory Module.
\newblock University of Utah, Tech. Rep., pp. 1--24, 2012.

\bibitem[Da95]{Davison95}
Davison, Andrew: Twelve Ways to Fool the Masses When Giving Performance Results
  on Parallel Computers.
\newblock pp. 38--42, 1995.

\bibitem[Da17]{conf/fpga/DaiHCXWY17}
Dai, Guohao; Huang, Tianhao; Chi, Yuze; Xu, Ningyi; Wang, Yu; Yang, Huazhong:
  {ForeGraph}: Exploring Large-scale Graph Processing on Multi-{FPGA}
  Architecture.
\newblock In: FPGA.
\newblock pp. 217--226, 2017.

\bibitem[Dr07]{Drepper07}
Drepper, Ulrich: What Every Programmer Should Know About Memory.
\newblock Red Hat, Inc, 11, 2007.

\bibitem[DRF20]{journals/corr/abs-2007-07595}
Dann, Jonas; Ritter, Daniel; Fröning, Holger: Non-Relational Databases on
  {FPGA}s: Survey, Design Decisions, Challenges.
\newblock CoRR, abs/2007.07595, 2020.

\bibitem[Gh19]{journals/pomacs/GhoseLHCM19}
Ghose, Saugata; Li, Tianshi; Hajinazar, Nastaran; Cali, Damla~Senol; Mutlu,
  Onur: Demystifying Complex Workload-DRAM Interactions: An Experimental Study.
\newblock Proc. {ACM} Meas. Anal. Comput. Syst., 3(3):60:1--60:50, 2019.

\bibitem[Ha16]{conf/micro/HamWSSM16}
Ham, Tae~Jun; Wu, Lisa; Sundaram, Narayanan; Satish, Nadathur; Martonosi,
  Margaret: Graphicionado: A High-Performance and Energy-Efficient Accelerator
  for Graph Analytics.
\newblock In: MICRO.
\newblock pp. 56:1--56:13, 2016.

\bibitem[HB15]{conf/sc/HoeflerB15}
Hoefler, Torsten; Belli, Roberto: Scientific Benchmarking of Parallel Computing
  Systems: Twelve Ways to Tell the Masses When Reporting Performance Results.
\newblock In: {SC}.
\newblock pp. 73:1--73:12, 2015.

\bibitem[KYM16]{journals/cal/KimYM16}
Kim, Yoongu; Yang, Weikun; Mutlu, Onur: Ramulator: A Fast and Extensible {DRAM}
  Simulator.
\newblock {IEEE} Comput. Archit. Lett., 15(1):45--49, 2016.

\bibitem[LK14]{LeskovecK14}
Leskovec, Jure; Krevl, Andrej: , {SNAP Datasets}: {Stanford} Large Network
  Dataset Collection.
\newblock \url{http://snap.stanford.edu/data}, June 2014.

\bibitem[Lu07]{LumsdaineGHB07}
Lumsdaine, Andrew; Gregor, Douglas; Hendrickson, Bruce; Berry, Jonathan:
  Challenges in Parallel Graph Processing.
\newblock Parallel Processing Letters, 17(01):5--20, 2007.

\bibitem[MBK02]{conf/vldb/ManegoldBK02}
Manegold, Stefan; Boncz, Peter~A.; Kersten, Martin~L.: Generic Database Cost
  Models for Hierarchical Memory Systems.
\newblock In: PVLDB.
\newblock pp. 191--202, 2002.

\bibitem[RCJ11]{journals/cal/RosenfeldCJ11}
Rosenfeld, Paul; Cooper{-}Balis, Elliott; Jacob, Bruce: DRAMSim2: {A} Cycle
  Accurate Memory System Simulator.
\newblock {IEEE} Comput. Archit. Lett., 10(1):16--19, 2011.

\bibitem[Xu17]{conf/icws/XuZLSGWLZ17}
Xu, Chongchong; Zhou, Jinhong; Lu, Yuntao; Sun, Fan; Gong, Lei; Wang, Chao; Li,
  Xi; Zhou, Xuehai: Evaluation and Trade-offs of Graph Processing for Cloud
  Services.
\newblock In: {IEEE} {ICWS}.
\newblock pp. 420--427, 2017.

\bibitem[Ya18]{conf/IEEEpact/Yao0L0H18}
Yao, Pengcheng; Zheng, Long; Liao, Xiaofei; Jin, Hai; He, Bingsheng: An
  Efficient Graph Accelerator with Parallel Data Conflict Management.
\newblock In: {PACT}.
\newblock pp. 8:1--8:12, 2018.

\bibitem[Ya19]{conf/islped/YanHLALMDYZF019}
Yan, Mingyu; Hu, Xing; Li, Shuangchen; Akgun, Itir; Li, Han; Ma, Xin; Deng,
  Lei; Ye, Xiaochun; Zhang, Zhimin; Fan, Dongrui; Xie, Yuan: Balancing Memory
  Accesses for Energy-Efficient Graph Analytics Accelerators.
\newblock In: ISLPED.
\newblock pp. 1--6, 2019.

\bibitem[ZCP15]{conf/reconfig/ZhouCP15}
Zhou, Shijie; Chelmis, Charalampos; Prasanna, Viktor~K.: Optimizing memory
  performance for {FPGA} implementation of {PageRank}.
\newblock In: ReConfig.
\newblock pp. 1--6, 2015.

\bibitem[Zh19]{journals/tpds/ZhouKPSW19}
Zhou, Shijie; Kannan, Rajgopal; Prasanna, Viktor~K.; Seetharaman, Guna; Wu,
  Qing: {HitGraph:} High-throughput Graph Processing Framework on {FPGA}.
\newblock {IEEE} Trans. Parallel Distrib. Syst., 30(10):2249--2264, 2019.

\bibitem[ZL18]{conf/fpga/ZhangL18}
Zhang, Jialiang; Li, Jing: Degree-aware Hybrid Graph Traversal on {FPGA-HMC}
  Platform.
\newblock In: FPGA.
\newblock pp. 229--238, 2018.

\end{thebibliography}

\newpage
\appendix
\section{Appendix}
\label{sec:appendix}

\begin{table}[ht]
\centering
\scriptsize
\begin{tabular}{l l r r r r r r r}
     & & \multicolumn{7}{c}{Graph} \\
    \cline{3-9}
    Problem & Measurement & bk & wt & rd & lj & tw & r21 & r24 \\
    \hline
    \multirow{2}{*}{SpMV} & Ground truth & 0.0032 & 0.0050 & 0.0028 & 0.0362 & 0.6525 & 0.0567 & 0.1435 \\
     & Simulation & 0.0026 & 0.0066 & 0.0027 & 0.0411 & 0.8184 & 0.0484 & 0.0770 \\
    \hline
    \multirow{2}{*}{PR} & Ground truth & 0.0030 & 0.0045 & 0.0027 & 0.0327 & 0.5904 & 0.0534 & 0.1403 \\
     & Simulation & 0.0026 & 0.0066 & 0.0027 & 0.0411 & 0.8184 & 0.0484 & 0.0770 \\ 
    \hline
    \multirow{2}{*}{SSSP} & Ground truth & 0.7824 & 0.0255 & 1.1133 & 0.5921 & 5.5768 & 0.9671 & 0.9213 \\
     & Simulation & 1.2554 & 0.0027 & 1.3436 & 0.3872 & 6.2380 & 0.0725 & 0.1111 \\
    \hline
    \multirow{2}{*}{WCC} & Ground truth & 1.7690 & 0.0460 & 1.4800 & 0.4130 & 6.6170 & 0.4500 & 1.1080 \\
     & Simulation & 1.8578 & 0.0461 & 1.4526 & 0.4694 & 9.4139 & 0.4653 & 0.9307 \\
\end{tabular}
\caption{HitGraph measurements in seconds}
\label{tab:hitgraph}
\end{table}

\vspace{.3cm}

\begin{table}[ht]
\centering
\scriptsize
\begin{tabular}{l l r r r r r r}
     & & \multicolumn{6}{c}{Graph} \\
    \cline{3-8}
    Problem & Measurement & sd & db & yt & wt & lj & or \\
    \hline
    \multirow{2}{*}{BFS} & Ground truth & 2.867 & 2.397 & 1.899 & 1.653 & 3.370 & 3.638 \\
     & Simulation & 2.880 & 2.515 & 2.530 & 1.999 & 2.946 & 3.192 \\ 
    \hline
    \multirow{2}{*}{PR} & Ground truth & 2.242 & 1.931 & 1.560 & 1.318 & 1.921 & 2.587 \\
     & Simulation & 2.518 & 1.944 & 1.978 & 1.283 & 1.926 & 2.920 \\ 
    \hline
    \multirow{2}{*}{WCC} & Ground truth & 2.950 & 2.468 & 1.954 & 1.729 & 2.407 & 3.365 \\
     & Simulation & 2.634 & 2.183 & 2.284 & 1.532 & 2.254 & 2.998 \\
\end{tabular}
\caption{AccuGraph measurements in GREPS}
\label{tab:accugraph}
\end{table}

\end{document}